\documentclass[preprint,12pt]{elsarticle}

\usepackage{amssymb}
\usepackage{amsmath}

\usepackage[utf8]{inputenc}

\usepackage{siunitx}
\DeclareSIUnit\townsend{Td}

\usepackage{hyperref}

\usepackage{tabularx}
\usepackage[draft,margin]{fixme}
\fxsetup{layout=pdfcmargin}

\usepackage{cleveref}
\usepackage{subcaption}
\usepackage{xcolor}

\usepackage{fancyvrb}

\usepackage{listings}

\lstdefinestyle{mystyle}{
	basicstyle=\ttfamily\scriptsize,
	breakatwhitespace=false,         
	breaklines=true,                 
	captionpos=b,                    
	keepspaces=true,                                   
	showspaces=false,                
	showstringspaces=false,
	showtabs=false,                  
	tabsize=2,
	frame = single, 
	backgroundcolor = \color[RGB]{255,255,235}
}

\newcounter{bla}

\journal{Computer Physics Communications}

\begin{document}
	
	\begin{frontmatter}
		
		
		
		\title{Betaboltz: a Monte-Carlo simulation tool for gas scattering processes}
		
		\author[a]{M. Renda\corref{author}}
		\author[a,b]{D. A. Ciubotaru}
		\author[c]{C. I. Banu}
		
		\cortext[author] {Corresponding author.\\\textit{E-mail address:} michele.renda@cern.ch}
		\address[a]{IFIN-HH, Particles Physics Department, M\u{a}gurele, Romania}
		\address[b]{Faculty of Physics, University of Bucharest, Bucharest - Măgurele, Romania}
		\address[c]{Faculty of Electronics, Telecommunications and Information Technology, University Politehnica of Bucharest, Romania}
		
		\begin{abstract}
			We present an open-source code for the simulation of electron and ion transport for user-defined gas mixtures with static uniform electric and magnetic fields. The program provides microscopic interaction simulation and is interfaced with cross-section tables published by LXCat[1]. The framework was validated against drift velocity tables available in literature obtaining an acceptable match for atomic and non-polar molecular gases with spherical symmetry. The code is written in \texttt{C++17} and is available as a shared library for easy integration into other simulation applications. 
		\end{abstract}
		
		
	\end{frontmatter}
	
	
	
	{\bf PROGRAM SUMMARY}
	
	\begin{small}
		\noindent
		{\em Program Title: Betaboltz}                                \\
		{\em Licensing provisions: LGPL v3}                              \\
		{\em Programming language: C++17}                               \\
		\noindent
		{\em Nature of problem:}
		Simulations of electron and ion transport in arbitrary gas mixture under static uniform electric and magnetic fields.\\
		{\em Solution method:}
		Particle motion using classical and relativistic equation via interaction sampling using Monte-Carlo techniques. \\
		{\em Additional comments including Restrictions and Unusual features:} At the time of writing only static uniform electromagnetic fields are supported. However, the implementation of arbitrary fields can be added given an analytical solution is available. A custom XML format for cross-section was developed, because full compatibility with LXCat [1] XML format was not possible. Cross-section databases in the new format are available in the download section of the LXCat site [2].

	\end{small}

	\section{Introduction} \label{sec:introduction}

A comprehensive understanding of the charged particle transport in low-temperature gas mixtures, has crucial importance for the design and simulation of gas-based radiation detectors. The modeling of these processes can be performed by numerically solving the Boltzmann equation \cite{boltzmann1896vorlesungen,ness1994multi}, or by sampling the motion of charged particles using Monte-Carlo techniques \cite{biagi1999monte}. Both approaches require the knowledge of the possible interaction types for all the components of the gas mixture. This information can be obtained using theoretical quantum models or by measurements of the electron/ion swarm experiments.

When performing any calculation involving particle transport in low-temperature gases, we have to choose which cross-section databases to use in the calculations. While, in the past, these tables had to be searched within the extensive available literature, now it is possible to access publicly available database such as LXCat \cite{pitchford2017lxcat}, collecting cross-section tables, drift velocities, and other swarm attributes, for an extensive set of gases.

Several Boltzmann and Monte-Carlo simulation frameworks exist, but none of them can perform full detector simulations or have features like multi-thread execution. A well-known and widespread solution is \texttt{Magboltz} \cite{magboltz}, an electron drift solver which contains the cross-section tables calculated by Biagi et al. \cite{biagi1999monte} and allows fast determination of the electron swarm properties for a reasonable set of gases. In the same class of solvers, we can mention \texttt{BOLSIG+} \cite{bolsig}, a well-known Boltzmann solver and \texttt{METHES} \cite{rabie2016methes}, which allows Monte-Carlo simulations for an arbitrary static electric field using cross-sections tables available from LXCat. In addition to the previous solutions, we should mention \texttt{PyBoltz} \cite{alatoum2020electron}, a \texttt{Cython} porting of the original \texttt{Magboltz} code and \texttt{pyMETHES} \cite{pymethes}, a \texttt{Python} port of the \texttt{METHES} package.

In this paper, we present a flexible and robust solution for the simulation of particle drift in a custom detector's using Monte Carlo methods. We will present here a list of the main characteristics of the simulation tool that we have developed:
\vspace{-0.1cm} 
\begin{itemize}
	\setlength\itemsep{0pt}
	\item \texttt{C++} shared library interface-able with existing codes.
	\item Multi-thread execution.
	\item Access to cross-section tables from LXCat.
	\item Possibility to simulate detector composed by multiple chambers with different electromagnetic fields and gases.
	\item Supports arbitrary static uniform electromagnetic fields.
	\item Dimensional check at compile time.
\end{itemize}

Our program's most important feature is that we do not perform any calculation of the swarm attribute: the software will calculate the motion of every electron/ion inside the gas and will call a set of \textit{handlers} for every collision. The user will be able to create custom handlers and calculate the macroscopic quantities of interest (however, in \cref{sec:tutorial}, we describe a command-line utility that can be used to calculate the most common swarm attributes). This characteristic allows our solution to be used in various applications, ranging from the simulation of avalanche amplifications in detectors to the determination of drift velocities in gases under strong electromagnetic fields.

An important aspect is that we do not provide any recommended cross-section data, but we leave the end-user responsible for choosing the proper set of tables. Selecting the right table set can be a difficult task in some cases, especially when there are several cross-section databases with non-negligible differences, as shown by \cite{raju2004electron}. In addition, no correction or parametric fitting is performed by our solution, providing a result based on clear and definite formulas with no arbitrary correction factors.

This paper begins with a brief description of the theoretical framework, in \cref{sec:theory}. The library architecture is presented in \cref{sec:architecture,sec:multi_thread}, while \cref{sec:trial}, will present and discuss a new algorithm that can be used to choose an efficient trial frequency. In \cref{sec:tutorial}, we provide some brief examples of the usage of our framework.  In \cref{sec:benchmark}, we will test our solution against data available in literature, highlighting the performance and the limitations of our theoretical framework, which are further discussed in \cref{sec:limitation,sec:conclusion}.


\section{Theoretical framework} \label{sec:theory}

We will report here the theoretical framework used in our simulation tool. A charged particle, under the effect of an electric and magnetic field, will be subject to an acceleration. For non-relativist velocities, it is possible to calculate the evolution of the particle state using the classical equation of motions, as presented in our previous work \cite{renda2018monte}. For higher energies, relativistic formulas can be used, as shown by S. Chin \cite{chin2009relativistic}, providing more accurate results with the cost of increased computation time.

The collision occurs after a random time $t_c$, depending on the particle speed and the gas mixture combined cross-section \cite{milloy1977validity}:
\begin{align}
\ln\left(\frac{1}{1-R}\right)  = \int_{0}^{t_c} \nu\left(t\right) dt \label{eq:freetime}
\end{align}
where $R \in \left[0,1\right)$ is a uniform random number and $\nu(t)$ is the interaction frequency of the particle in the gas. The value of the interaction frequency depends on the particle energy and the gas composition. For a gas mixture of $N$ components, where each component has an elastic interaction mode and $J$ inelastic ones, we have:
\begin{align}
\nu\left(\varepsilon\right) = \sqrt{\frac{2 \varepsilon}{m}} \sum_{k=1}^{N} n_k \left(\sigma_k\left(\varepsilon\right) + \sum_{i=1}^{J_k} \sigma_{ki} \left(\varepsilon\right)\right)  \label{eq:frequency}
\end{align}
where $\varepsilon$ and $m$ are, respectively, the particle energy and mass, $n_k$ is the molecule number density and $\sigma_k$ and $\sigma_{ki}$ the elastic and inelastic cross-sections.

Resolving \cref{eq:freetime} for every interaction would require the numerical solution of the integral for each collision, thus requiring a substantial computing time. To avoid that, it was used the \textit{null collision} technique presented by Skullerud \cite{skullerud1968stochastic}: this technique allows to bypass the numeric integration of $\nu(t)$ replacing it with a constant trial frequency $\nu' \geqslant \nu(t)$. The \cref{eq:freetime} now becomes:
\begin{align}
t_c = - \frac{\ln(1-R)}{\nu^\prime} \label{eq:random_time}
\end{align}

This substitution is possible only if we consider a fraction of the interaction as \textit{null collision}, interactions that does not alter the direction and energy of the particle. To decide if an interaction should be considered \textit{null}, we use a uniform random number $R \in [0,1)$ and mark as \textit{null} all collisions satisfying the condition:
\begin{align}
R > \nu(t) / \nu' \label{eq:null_fraction}
\end{align}
where $\nu(t)$ is the real interaction frequency right before the next collision.

It is important to remark that determining a reasonable value of $\nu'$ is quite important because it directly affects computing performance. A too high value will generate a high number of \textit{null collisions}, decreasing the computing performance. A too low value will generate situations where $\nu(t) > \nu'$, invalidating the result and requiring a re-computation of the step. Because several strategies are available \cite{skullerud1968stochastic,brennan1991optimization} to compute a proper value for the trial frequency, we decided to allow the user to choose the best strategy to determinate the $\nu'$ value (details in \cref{sec:trial}).

Further, we have to handle the collisions: for a correct calculation, we would need the differential cross-sections for each process, in the form of $\sigma(\varepsilon, \theta)$. Unfortunately, such cross-section tables are difficult to be obtained experimentally, and in the literature can be found such tables just for a limited set of gases. Instead, it is quite common to find integral cross-section for all gases used in gas detectors in the form of total elastic cross-section $\sigma_{el}(\varepsilon)$ or momentum transfer cross-sections $\sigma_{mt}(\varepsilon)$.

In our implementation we decided to use the approach presented by Okhrimovskyy et. al. \cite{okhrimovskyy2002electron} (however, we have to mention that there are other valid approaches, such as the one presented by Longo and Capitelli \cite{longo1994simple}). This approach allows using a pseudo-differential cross-section generated by the combination of both $\sigma_{el}$ and $\sigma_{mt}$.  In this way, the scattering angle $\theta$ for atomic gases like $Ar$, $Ne$, $Xe$, etc. can be calculated using a formula based on the screened Coulomb potential:
\begin{align}
\cos(\theta) = 1 - \dfrac{2 R}{1 + 8 \; \varepsilon \; (1 - R)} \label{eq:collision_angle}
\end{align}
where $R \in [0,1)$ is a uniform random number and $\varepsilon$ is the dimension-less energy in atomic units. For non-polar molecular gases, like $CO_2$, $CH_4$ or $O_2$, a similar formula can be used \cite{okhrimovskyy2002electron}:
\begin{align}
\cos(\theta)=1-\dfrac{2 R(1-\xi)}{1+\xi(1-2 R)}  \label{eq:collision_angle_xi}
\end{align}
where $\xi$ is a dimensionless value derived solving the equation:
\begin{align}
\frac{\sigma_{mt}(\varepsilon)}{\sigma_{el}(\varepsilon)}=\dfrac{1-\xi}{2 \; \xi^{2}}\left((1+\xi) \ln \dfrac{1+\xi}{1-\xi}-2 \xi \right)
\end{align}

Finally, we calculate the energy exchange for a given interaction, knowing the deviation angle $\theta$. We use the model of Fraser and Mathieson \cite{fraser1987monte}, which gives an analytical expression for both elastic collisions:
\begin{align}
\dfrac{\varepsilon_f}{\varepsilon_i}  = 1 - \dfrac{2  m M (1 - \cos \theta)}{(m+M)^2}  \label{eq:delta_energy_ela}
\end{align}
and inelastic collision:
\begin{align}
\dfrac{\varepsilon_f}{\varepsilon_i}  = 1 - \dfrac{M}{m + M} \dfrac{\varepsilon_k}{\varepsilon_i} + \dfrac{2 m M}{(m + M)^2} \left(\sqrt{1 - \dfrac{m+M}{M} \dfrac{\varepsilon_k}{\varepsilon_i}} \cos \theta - 1 \right) \label{eq:delta_energy_ine}
\end{align}
where $m$ is the incoming particle mass and $M$ is the target molecule mass, $\varepsilon_i$ and $\varepsilon_f$ is the energy before and after the collision and $\varepsilon_k$ is the threshold energy characteristic of the given inelastic process.

\Cref{eq:delta_energy_ine} can be used to calculate the energy transfer for any inelastic process. However, there are two exceptions: for attachment processes, where the $\varepsilon_k$ can be zero, we can assume all the energy is transferred, considering them as plastic collisions. In the case of ionizations, the equations remain valid. However, another step is necessary: after the extraction of the electron, we have to calculate the energy transfer between the bullet and the extracted electron. We can model it as an elastic collision. Failing to do so, may cause the drift velocity to diverge for high fields, when the ionization processes overtake the elastic ones.

\subsection{Effect of gas temperature}

\Cref{eq:delta_energy_ela,eq:delta_energy_ine} are very useful to calculate the energy transfer at each collision, both for elastic and inelastic collisions. However, analyzing these formulas, we can notice that, while we account for the mass of both bullet particle and target molecule, we do assume the target at rest. This is a very reasonable assumption in any electron/ion drift experiment: at normal conditions the mean energy of any gas molecule is in the scale of $\approx \SI{25}{\milli\electronvolt}$. This is much lower of the mean energies of electrons/ions in a common drift experiment, which is in the order of tens electron-volts.

However, we want our solution to be suitable for a wide range of EM fields, with values ranging between \SI{1e-3}{\townsend} and up to over \SI{1e4}{\townsend}. For values lower than \SI{0.01}{\townsend}, we can not assume anymore the gas components at rest. Each molecule will have a speed distributed according to the Maxwell-Boltzmann distribution which will affect the charged particle drift in the gas.

We calculate, in \cite{renda2020effects}, the effects of the temperature of the gas in the collision dynamics: 
\begin{align}
\textbf{V}_1' &= \textbf{V}_{cm} +   \sqrt{| \textbf{V}_1 - \textbf{V}_{cm} |^2 - \frac{2 \varepsilon_k}{m + M} \left(\frac{M}{m}\right)} \; \hat{\textbf{u}}_1' \\
\textbf{V}_2' &= \textbf{V}_{cm} +   \sqrt{| \textbf{V}_2 - \textbf{V}_{cm} |^2 - \frac{2 \varepsilon_k}{m + M} \left(\frac{m}{M}\right)} \; \hat{\textbf{u}}_2'
\end{align}
where $\textbf{V}_{1}$, $\textbf{V}_{2}$ and $\textbf{V}_{cm}$ are, respectively the bullet particle, target and the center-of-momentum velocities in the laboratory frame, and $\hat{\textbf{u}}_1'$ and  $\hat{\textbf{u}}_2'$ the new bullet and target velocities in the center-of-momentum frame.  

\subsection{Extension of the cross-section tables}

The extended range of the EM fields that we decided to support, put stress on the cross-section tables we can use. Many tables have a limited energy range being limited, usually, to $\approx \SI{1}{\kilo\electronvolt}$. This could seem a reasonable value, but we have not to ignore the fact that, if we have a big number of bullets and/or interactions, a fraction of them will overpass this energy boundary. Fortunately, if we look at a log-log plot of the cross-section tables, we can realize that it is possible to easily extend these tables.

For elastic processes, we can extend to the left (lower energies), using a constant value based on the first values of the table. For inelastic tables, instead, we can truncate the value to zero for values below the threshold energy. The right extension, for higher energies, we found it is possible to fit a straight line, using two near points:
\begin{align}
K &= \frac{\ln(\sigma_2 / \sigma_1)}{\ln(\varepsilon_2 / \varepsilon_1)} \\
A &= \sigma_2 / \varepsilon^K_2 = \sigma_1 / \varepsilon^K_1\\
\sigma(\varepsilon) &= A \; \varepsilon^K
\end{align}
where $K$ is a dimensionless number, $\sigma_1$ and $\sigma_2$ are the cross-sections in two points near the boundary, $\varepsilon_1$ and $\varepsilon_2$ are the respective energies and $\varepsilon$ is the requested energy.

\subsection{Determination of swarm attributes}

As stated in the introduction, our core library, \texttt{Betaboltz}, does not perform, by default, any computation of swarm attributes (drift velocity, ionization coefficients, diffusion, etc.), delegating to the final user the duty to create its own analysis using the provided handler mechanism. However, we can not ignore the importance of such attributes within the low temperature plasma community, so we provide a command-line utility, \texttt{drifter}, which allows their calculation without setting up a full analysis.

In this section, we will briefly present how such attributes are calculated. First, we use an infinite volume with a uniform static electric field. The axes are arranged so that the $z$-axis is aligned with the same direction of the electric field. A magnetic field may be specified at an arbitrary angle $\theta$ in the plane $xz$.

Then, $N$ primary electrons are placed at $(0,0,0)$ and they are left drifting for an $I$ number of real collisions. We then can repeat the simulation for $E$ events and, eventually, for $R$ runs (we adopt the division in events and runs as commonly used in the HEP community).

At the end of each event, we calculate the position of the individual particle along the z-axis. This information allows determining the drift velocity $W$, along the $z$-axis, using the following formula:
\begin{align}
	 W = \frac{1}{N} \sum^N_n \frac{z_n}{t_n}
\end{align}
where $z_n$ is the $z-$position of the $n$ particle when the particle is destroyed (because it reached the $I$ collisions count), and $t_n$ is the time when the destruction occurs. 

For the calculation of the Townsend coefficients, $\alpha$ and $\eta$, we use the concept of \textit{dummy} ionizations and attachments. We do calculate the energy exchange of such interactions, but we do not create/suppress any particle, limiting our-self to count the number of particles that would have been created or destroyed. This method is effective and allows to quickly calculate the coefficients in very high fields, where a full simulation would be hard to achieve::
\begin{align}
\left<\alpha\right> &= \frac{1}{N} \sum^N_n \frac{G_n}{z_n} \\
\left<\eta\right>   &= \frac{1}{N} \sum^N_n \frac{D_n}{z_n}
\end{align}
where $G_n$ is the number of new particles created by ionizations and $D_n$ the number of particles destroyed by attachments.
 
For the mean energy, we use a different approach. We perform a plain mean of the energies just before and after the collisions for all particles and events:
\begin{align}
\left<\varepsilon \right> &= \frac{1}{E N I} \sum^{E,N,I}_{e,n,i} \varepsilon_{n e}(t_i) \\
\left<\varepsilon'\right> &= \frac{1}{E N I} \sum^{E,N,I}_{e,n,i} \varepsilon_{n e}'(t_i)
\end{align}
where $n$ is the particle index, $e$ is the event index and $t_i$ is the time when the $i$ collision occurs.

For diffusion coefficients and magnitude, we calculate, at the event end, the mean velocity $W$ of the swarm along the $z$ axis:
\begin{align}
W &= \frac{1}{N} \sum^N_n z_n / t_n 
\end{align}
where $z_n$ is the position of the $n$ particle along the $z$ axis, and $t_n$ the time needed to move such distance. Then, for each particle, we calculate the displacement vector as:
\begin{align}
\textbf{d}_n &= \textbf{x}_n - t_n \; W \; \hat{\textbf{z}}
\end{align}
where $W$ is the mean velocity calculated above, $t_n$ the time when the particle is destroyed and $\hat{\textbf{z}}$ a dimensionless versor pointed along the $z$ direction. We can now calculate the transversal, $D_T$, and longitudinal, $D_L$, diffusion coefficients as:
\begin{align}
D_{\;\;}  &=  \frac{1}{N} \sum^N_n \frac{\left|\textbf{d}_n \right|^2}{t_n } \\
D_L       &=  \frac{1}{N} \sum^N_n \frac{\left(\textbf{d}_n \cdot \hat{\textbf{z}}\right)^2}{2 t_n } \\
D_T       &=  \frac{1}{N} \sum^N_n \frac{\left|\textbf{d}_n - d_n \cdot \hat{\textbf{z}}\right|^2}{4 t_n }
\end{align}
and their respective magnitude:
\begin{align}
\sigma_{\;\;} &=  \frac{1}{N} \sum^N_n 2 \left|\textbf{d}_n\right| \\
\sigma_L      &=  \frac{1}{N} \sum^N_n 2 \left|\textbf{d}_n \cdot \hat{\textbf{z}}\right| \\
\sigma_T      &=  \frac{1}{N} \sum^N_n 2 \left|\textbf{d}_n - d_n \cdot \hat{\textbf{z}}\right|
\end{align}


\section{Software architecture} \label{sec:architecture}

In this section, we will present the architecture of our software. The project is divided into several modules:

\begin{itemize}
	\item \texttt{Univec}: a custom-created library which allow vector operations with compile time dimensional analysis \cite{renda_univec_2018}.
	\item \texttt{ZCross}: a library that allows reading the cross-section tables in XML format \cite{renda_zcross_2018,renda2019zcross}.
	\item \texttt{ZCrossGPU}: an optional library that allows performing the cross-section interpolations on GPU \cite{banu_zcross_2019}.
	\item \texttt{Betaboltz}: the main simulation library \cite{renda_betaboltz_2018}.
	\item \texttt{Drifter}: a command-line utility to calculate the swarm  attributes of a gas mixture \cite{renda_drifter_2019}.
\end{itemize}

\begin{figure}
	\centering
	\includegraphics[width=0.85\columnwidth]{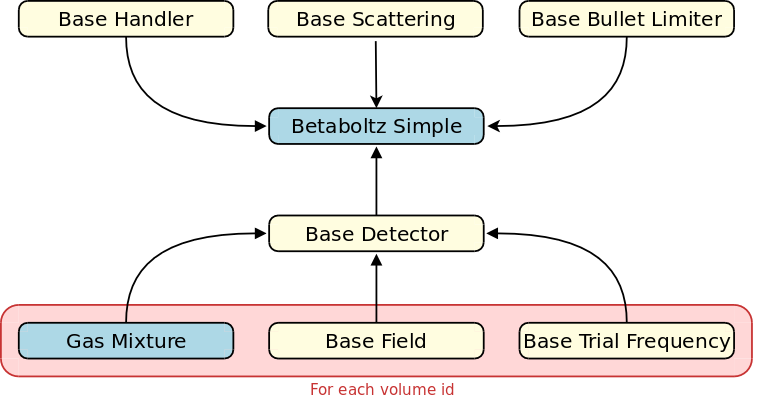}
	\caption[Layout of the software architecture]{In this schema we can find the relation between the classes of our library. The classes in yellow are abstract (we provide some implementations for the most common cases). Arrows represent usage, not inheritance.}
	\label{img:diagram}
\end{figure}

In \cref{img:diagram}, we show the organization of the classes in the \texttt{Betaboltz} software library. Some classes are defined as abstract classes, and we encourage the end-user to implement them for their specific needs, although we provide a set of concrete implementations to cover the most common scenarios. 

In this section, we will analyze each class, providing some details about their purpose and the implementations already existing by default. We want to remark that this is only a very brief introduction and more details and examples can be found in the software user guide \cite{betaboltz_user_guide}.

It is worth mentioning that we heavily use the concept of static dimensional checking \cite{schabel2003boost} aiming to reduce of the probability of software bugs and an improved type consistency. 

\subsection{Betaboltz Simple}

This class is the starting point of the library and allows the simulation of an arbitrary number of electrons or ions for an indefinite amount of time. To be able to perform its job, we need to provide:

\begin{itemize}
	\item A \texttt{BaseDetector}, providing the geometry, fields, and gases of the detector.
	\item A list of \texttt{BaseBulletLimiter}, providing the conditions which will halt the simulation.
	\item A list of \texttt{BaseHandler}, providing the actions to perform when something notable happens in the simulation.
\end{itemize}

The only mandatory element is \texttt{BaseDetector}, which divides the space into volumes with specific gas conditions and EM fields. The other two elements are optional: if no \texttt{BaseBulletLimiter} is added, a default one will be added, which will dispose the particles once outside the detector boundaries. If no \texttt{BaseHandler} is provided, no handler will be executed during the simulation.

\subsection{BaseBulletLimiter}

This class will take care to remove from the simulation particles when a specific condition occurs. It can be used to limit the simulation up to a particular time or a specific energy range. In \cref{tab:limiters}, is presented a list of existing implementations of this class, covering the most common usage. If none of the existing limiters satisfies the user’s needs,it is possible to create a custom limiter implementing the method \texttt{BaseBulletLimiter::isOver()}. 

\begin{table*}
	\centering
	\begin{tabularx}{\textwidth}{l X}
		\hline
		Class Name    &  Description \\ 
		\hline
		\rule{0pt}{20pt} \texttt{TimeBulletLimiter}    			&  Limit the simulation to a certain duration.\\ 
		\rule{0pt}{20pt} \texttt{DistanceBulletLimiter}    		&  Destroy any particle which goes too distant regards to a given point.\\ 
		\rule{0pt}{20pt} \texttt{EnergyBulletLimiter}    		&  Destroy all the particles when the energy goes outside the given energy range.\\ 
		\rule{0pt}{20pt} \texttt{ChildrenBulletLimiter}    		&  Limit the number of particles in the simulation \\ 
		\rule{0pt}{20pt} \texttt{InteractionBulletLimiter}    	&  Limit the number of interaction a particle can have in a simulation \\
		\rule{0pt}{20pt} \texttt{OutOfDetectorBulletLimiter}    &  Destroy any particle which goes outside the detector. This is the default if no other limiter is specified. \\
	\end{tabularx}
	\caption[Bullet limiters table]{This is the list of the classes extending the class \texttt{BaseBulletLimiter}. If you need a behavior not listed here, it is possible to create your own limiter extending the class \texttt{BaseBulletLimiter} and implementing the method \texttt{isOver()}.}
	\label{tab:limiters}
\end{table*}

\subsection{BaseHandler}

This class contains the actions to perform when notable events take place during the simulation. In \cref{tab:base-handler-methods}, are listed the methods which are called during the simulation. Two implementations of this class already exist: \texttt{PrintProgressHandler} is used to print to the console the simulation progress while \texttt{ExportCSVHandler} is used to write into a CSV file the result of a simulation. A custom handler can be created overriding one or more of the methods listed in the table mentioned above.

\begin{table}     
	\centering
	\begin{tabularx}{\columnwidth}{ l X }
		\hline
		Method      & Description \\ \hline
		\rule{0pt}{20pt} \texttt{onRunStart}                &    This method is called when the simulation starts.\\
		\rule{0pt}{20pt} \texttt{onRunEnd}                    &    This method is called when the simulation ends.\\
		\rule{0pt}{20pt} \texttt{onEventStart}                &    This method is called on the beginning of every event.\\
		\rule{0pt}{20pt} \texttt{onEventEnd}                &    This method is called at the end of every event.\\
		\rule{0pt}{20pt} \texttt{onBulletCreate}            &    This method is called when a new bullet is created (i.e. by an ionization).\\
		\rule{0pt}{20pt} \texttt{onBulletStep}                &    This method is called between two collisions. The 'from' prefix specify the state just after the last collision and the 'to' the state just before the current collision.\\
		\rule{0pt}{20pt} \texttt{onBulletCollision}            &    This method is called at every collision. The 'before' prefix specify the bullet state just before the collision, the 'after' just after. \\
		\rule{0pt}{20pt} \texttt{onBulletDestroy}            &    This method is called when a bullet is destroyed (i.e. by an attachment).\\
	\end{tabularx}
	\caption[Base handler methods]{Table representing the base handlers methods which are called during a simulation.}
	\label{tab:base-handler-methods}
\end{table}

\subsection{BaseDetector}  \label{sec:detector}

This class contains the detector geometry: the physical space is divided by volumes, identified by a non-negative integer. Each volume has its own \texttt{GasMixture}, \texttt{BaseField} and \texttt{BaseTrialFrequency}. Negative integers are reserved and used to indicate the out-of-detector state. Two implementations exist to perform a simulation without the need to define a detector: \texttt{InfiniteDetector}, a detector with a single infinite volume, and \texttt{BoxDetector}, a rectangular cuboid containing a single volume.

\subsection{GasMixture}

This class is used to define the gas mixtures. It is possible to define gases of any molecule, using common formula notations like \texttt{Si(CH3)4}: the software will compute the molecule mass and find a match in the cross-section database. Gas components can be specified by mass densities, molar densities or molecule number densities.

\subsection{BaseField}

This class specifies an electromagnetic field. For the moment, only uniform and static fields can be used. However, it can be easily expanded to non-uniform and non-static fields, given that an analytical solution can be implemented in the \texttt{moveParticle()} method. At this time, only two implementations are provided: \texttt{UniformFieldClassic}, which provides the classical motion equation for both $\textbf{E}$ and $\textbf{B}$ fields and the class \texttt{UniformFieldRelativisticChin} implementing the equation of motion for relativistic velocities as presented by S. A. Chin \cite{chin2009relativistic}.

\subsection{BaseScattering}

This class is used to define the scattering algorithm, i.e. how the $\theta$ and $\phi$ angle should be computed for each collision. While in our solution only two implementations were provided, isotropic and Okhrimovskyy's algorithm (described in \cref{sec:theory}), we decide to use a modular architecture, which allows adding and testing new scattering theories via, implementing the \texttt{BaseScattering::scatter()} method.

\subsection{BaseTrialFrequency}

This class provides the strategy used to determine the trial frequency during the simulation. This parameter is quite important because it directly affects the simulation performances. Some concrete implementations are discussed in \cref{sec:trial}.


\section{Multi-threading implementation} \label{sec:multi_thread}

As stated in \cref{sec:architecture}, we planned to have integrated support of multi-thread execution. Proper multi-thread implementation in \texttt{C++} is not a trivial task: if not carefully implemented, multi-thread execution can lead to run-time data races and deadlocks. This kind of errors, can block the simulation or, even worse, lead to incorrect results.

We decided to base our implementation on a common and well-tested solution named \texttt{OpenMP} \cite{dagum1998openmp}. This helped us to focus only on the logical implementation of the parallel sections, letting the library manage the data synchronization between the different threads.

\begin{figure}
	\includegraphics[width=0.80\columnwidth]{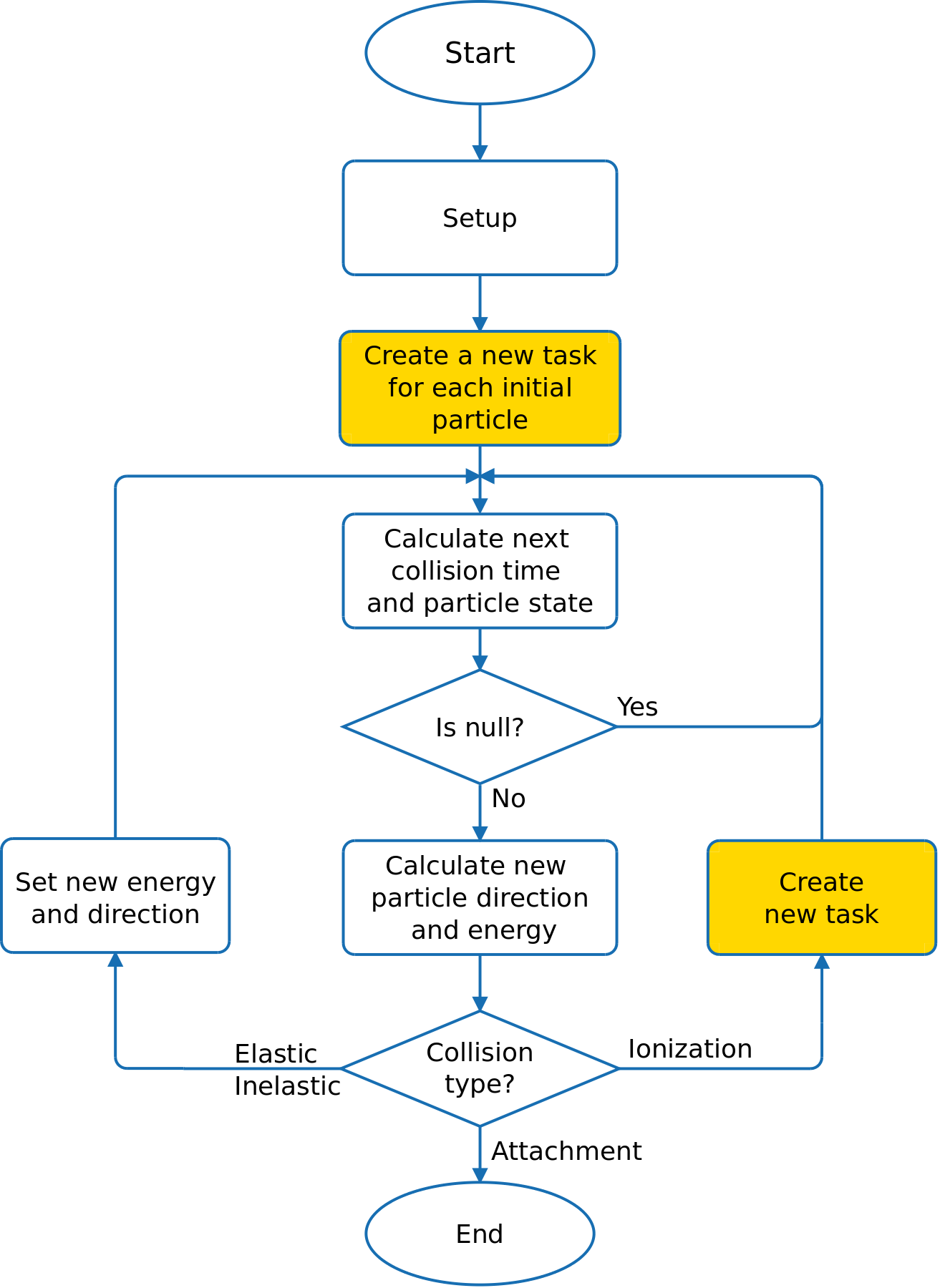}
	\centering
	\caption[Simulation flow chart]{Simplified flow chart representing the main steps taken during a simulation. Orange boxes represent points where a new parallel task is launched, when multi-threading support is enabled. Parallel tasks are executed concurrently until the maximum number of active threads is achieved.}
	\label{img:simulation_flowchart}
\end{figure}

As shown by \cref{img:simulation_flowchart}, we decided to use a separate task for each charged particle. Every new particle, created by an ionization, has its separate thread and can be executed concurrently with the other tasks. The \texttt{OpenMP} framework will schedule each task's execution and synchronize the shared data between the threads.

The drawback of this implementation is that only simulations containing multiple initial particles or presenting ionization processes can benefit from a higher number of execution threads.

Multi-threading support can be enabled or disabled via the \texttt{ENABLE\_OPENMP} cmake variable (default value is \texttt{ON}). The number of maximum active threads can be set using the common \texttt{OpenMP} methods (\texttt{omp\_set\_num\_threads()} function or the \texttt{OMP\_NUM\_THREADS} environment variable) or our \texttt{BetaboltzBase::setNumThreads()} function.

\section{GPU acceleration} \label{sec:gpu}

One of the most time-critical operations, during a simulation, is the solution of \cref{eq:frequency}. Solving this equation is computationally simple, requiring only basic arithmetic operations. However, if we analyze the real case usage of this equation, we realize it could become a critical bottle-neck, at least in some situations.

First, we have to sum all the components of the gas mixture (e.g. in common air, $\approx \num{10}$ components, to get a high precision simulation), when for each component, we have to sum all cross-section processes (between \numrange{10}{100} tables) and then, for each table, we have to perform a linear interpolation. This operation must be repeated for each collision, including the \textit{null} ones: we can realize how this operation can become very demanding when simulating complex gas mixtures.

However, due to the specific nature of this problem, we can take advantage of GPU computing to perform all the sum and interpolations using concurrent processing cores. C\u{a}lin Banu implemented this feature as an optional external library \cite{banu_zcross_2019}, which can be compiled and linked to the main \texttt{Betaboltz} framework. It uses the \texttt{CUDA} library to offload the calculation of the real frequency on the GPU.

This implementation should be considered a \textit{proof of concept} about the feasibility and the ability to provide results with the same accuracy of the non-GPU implementation. Additional work is required to improve the efficiency of the available solution and to make it competitive against the CPU only implementation.

\section{Trial frequency strategies} \label{sec:trial}

In \cref{sec:theory}, we highlighted the importance of choosing an optimized value for the trial frequency $\nu'$ used by the \textit{null collision} algorithm. In the original article by Skullerud \cite{skullerud1968stochastic}, an algorithm is discussed on how a proper trial frequency can be determined. During our preliminary testing, we realized how vital a proper trial frequency is to get good performances. Several studies focus their attention on optimizing the execution of Monte-Carlo code using different algorithms built on lookup tables based on particle energy \cite{brennan1991optimization}.

Let's now introduce the concept of simulation efficiency. To do so, let's briefly recall the \textit{null collision} algorithm: during a simulation, after a time related to $1/\nu'$ (see \cref{eq:random_time}) we will have a potential interaction, which can be \textit{real} or \textit{null}. Then, we use a uniform random number to decide is the collision is null or not, as shown by \cref{eq:null_fraction}. Also, we have to verify if, just before the collision, the criteria $\nu(\varepsilon) \le \nu'$ is respected: if not, we need to revert to the previous valid state and try again with an increased trial frequency. We want to remark that reverting to a previous state has a computational cost higher than a null collision, due to the need of restoring the previous state, discarding the null-collisions since the last real collision.

Observing a simulation, we can determinate three main figures related to the simulation frequency: the first one is the real frequency, the number of collisions per second altering the particle energy and direction, named $\nu_{real}$, the second one is the number of null collisions, $\nu_{null}$, and the last one, how many times the simulations had to be halted and restarted from a valid state, $\nu_{fail}$. We can define the efficiency of our simulation as:
\begin{align}
\eta = \dfrac{\nu_{real}}{\nu_{real} + \nu_{null} + \nu_{fail}}
\end{align}

It is important to notice that the chosen trial frequency is not constant, but we can change it during the simulation to adapt to the current particle energy. However, changing it at each collision will invalidate the null-collision algorithm, so a trade-off must be accepted to ensure the correctness of the result. We implemented this trade-off in the form of a \textit{grace} number of \textit{real} collisions, expressed as a number of real collisions.
\begin{enumerate}
	\item We query for the initial trial frequency, and we get a trial frequency and the number of \textit{grace} collisions.
	\item For each \textit{null} or \textit{real} collision, we keep constant the trial frequency, and we decrease the grace collisions counter.
	\item When the \textit{grace} counter reaches zero, we calculate a new trial frequency (using the \texttt{onNull} or \texttt{onReal} method according to the last collision type).
	\item When a collision fails, regardless of the \textit{grace} counter value, we restore the state of the last \textit{real} collision, and we query a new trial frequency using the \texttt{onFail} method, resetting the \textit{grace} counter.
\end{enumerate}

In the library we developed, we provide four concrete implementations of the \texttt{BaseTrialFrequency} strategy: two static strategies (the collision frequency does not change during the simulation) and two adaptive ones: we found a \num{5000} collisions grace period to be a reasonable value for our adaptive trial strategy.

\subsection{FixedTrialFrequency}

This is the simplest one, the user can choose the trial frequency and it will remain fixed for the entire simulation. If the frequency is too low at any point of the simulation, the execution will halt and an exception will be thrown. This strategy is recommended when we know the nature of the problem well, and we know the maximum frequency we can get in the energy range of interest.

\subsection{LazyTrialFrequency}

This is the safest algorithm available to determine the trial frequency, but unfortunately, it has quite a poor performance when applied to real cases. Given a gas mixture, it scans the local maximum and minimum of the collision frequencies of the gas components. For each local peak, the algorithm calculates $\nu(\varepsilon)$ and at the end will keep the highest value from those calculated.

Using the value calculated in this way, we can assert that $\nu_{fail} = 0$, so the simulation will never revert to a previous state. The drawback of this solution is that $\nu_{null} \gg \nu_{real}$, providing, in our testing, for several gases, has an efficiency low such as $\eta \approx 2.5\%$.

\subsection{BinnedTrialFrequency}

\begin{figure}
	\centering
	\includegraphics[width=.95\columnwidth]{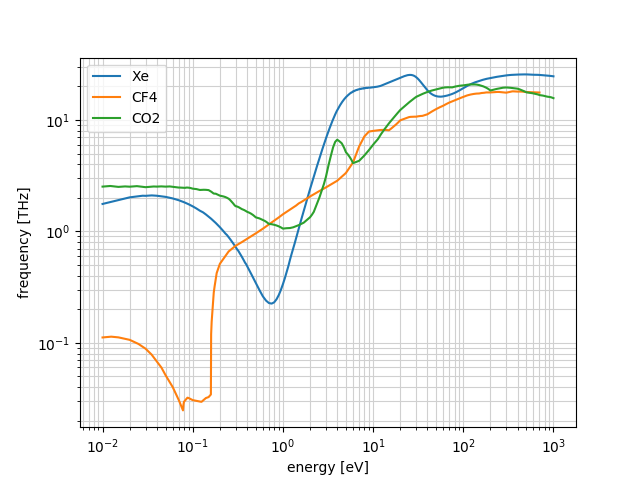}
	\caption{Real interaction frequencies between \SIrange{0.01}{1000}{\electronvolt} for electrons in pure $Xe$, $CF_4$ and $CO_2$. Cross-section data taken from \cite{cop2020lxcat,biagi2019lxcat,bordage2019lxcat,phelps2019lxcat}.}
	\label{fig:plot_frequencies}
\end{figure}

Conceptually similar to \texttt{LazyTrialFrequency}, it substantially improves its performance by dividing the energy range in bins. As we can see in \cref{fig:plot_frequencies}, the real interaction frequency can change abruptly within a small energy range. Using this class, the user may control the number of bins for each \textit{decade} (we define \SIrange[range-phrase=--]{1}{10}{\electronvolt} as a \textit{decade}, \SIrange[range-phrase=--]{10}{100}{\electronvolt} as another, and so on). For each bin, we calculate the maximum possible frequency achievable. When we have a fail, e.g. when we cross from a bin to another, we retry with the maximum frequency between the two bins. If it still fails, then we throw an exception.

\subsection{VariableTrialFrequency}

\begin{figure}
	\centering
	\includegraphics[width=.95\columnwidth]{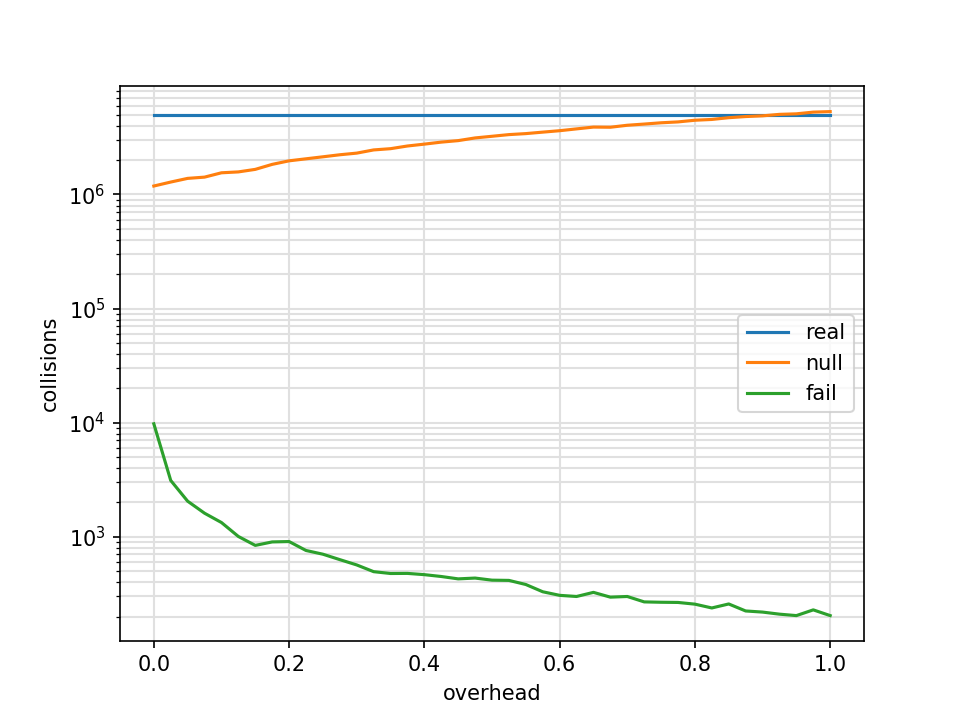}
	\caption[Variable trial frequencies]{Proportion between real, null and fail collisions during a simulation for different values of the parameter $\zeta$ for $CO_2$ at a reduced electric field of \SI{15}{\townsend}. The simulation was halted at \num{250000} real collisions. Cross-section data from \cite{biagi2019lxcat,phelps2019lxcat}.}
	\label{fig:plot_efficiency_zeta}
\end{figure}

\begin{figure}
	\centering
	\includegraphics[width=.95\columnwidth]{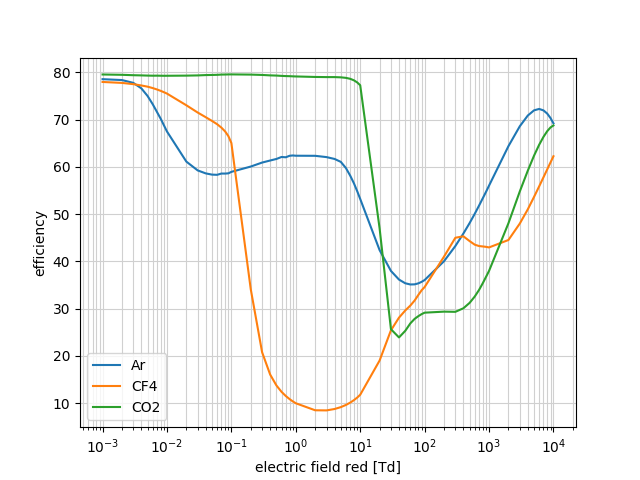}
	\caption{Simulation efficiency for variable trial frequency ($\zeta = \SI{25}{\percent}$) for $Ar$, $CF_4$ and $CO_2$. Cross-section data from \cite{biagi2019lxcat,phelps2019lxcat,bordage2019lxcat,bsr2020lxcat}.}
	\label{fig:plot_efficiency_gases}
\end{figure}

During the development of the previous strategies, we noticed that the real interaction frequency is strictly related to the particle energy and gas conditions. This algorithm will try to balance the $\nu_{null}$ and $\nu_{fail}$, to achieve a good efficiency $\eta$. If we have $\nu'=\nu_{real}$, we will surely get $\nu_{null} = 0$ but, from the other side, we can get a $\nu_{fail} \gg \nu_{real}$, with the related performance loss. In this algorithm, we set, at the end of each \textit{grace} period, a new value for $\nu'$ with a specified overhead $\zeta$ assigned by the user (default value \SI{25}{\percent}):
\begin{align}
\nu' = \left(1 + \zeta \right)\cdot \nu(\varepsilon)
\end{align}
Using this approach we were able to obtain an efficiency $\eta$ up to $\approx \SI{80}{\percent}$ in several simulations when using a $\zeta \approx \SI{25}{\percent}$, as shown in \cref{fig:plot_efficiency_zeta,fig:plot_efficiency_gases}. It is important to remark that this is a state-less approach: we do not keep any state variable for the moving particles to reduce memory usage. State-full approaches could improve the efficiency to higher values, but it was not the main aim of our study.

\subsection{Writing custom trial frequency strategies}
It is possible to write a custom trial frequency strategy, to increase the efficiency in certain situations. A custom strategy can be written by extending the class \texttt{BaseTrialFrequency}, implementing these methods: 
\begin{itemize}
	\item \texttt{getInitialTrialFrequency}
	\item \texttt{getNextGrace}
	\item \texttt{getNextTrialFrequencyOnReal}
	\item \texttt{getNextTrialFrequencyOnNull}
	\item \texttt{getNextTrialFrequencyOnFail}
\end{itemize}

It is important to notice that these methods are \texttt{const}, forcing to write a state-less algorithm. In the future, we will evaluate the opportunity to make these methods \texttt{non-const}, allowing a more efficient strategy at the cost of increased memory usage. During the simulation, it is possible to check the current efficiency using the method \texttt{BetaboltzSimple::getStatsEfficency()}.

\section{Installation and Usage} \label{sec:tutorial}

In this section, we will briefly present some helpful information about the installation and the usage of the framework we developed. First, we need to install \texttt{ZCross}. We can get it via its public repository:
\begin{lstlisting}[language=Bash]
$ git clone https://gitlab.com/micrenda/zcross.git
$ mkdir zcross/build && cd zcross/build
$ cmake -DCMAKE_INSTALL_PREFIX=/opt/zcross ..
$ make && sudo make install
\end{lstlisting}

Then we need to download the cross-section databases of interests. It is possible to download them from the LXCat download page \cite{lxcat2021offline}. We recommend saving into a personal directory and setting an environment variable pointing there:
\begin{lstlisting}[language=Bash]
$ mkdir ~/.zcross_data/
$ export ZCROSS_DATA=~/.zcross_data/
\end{lstlisting}

Finally, it is possible to test the \texttt{ZCross} installation:
\begin{lstlisting}[language=Bash]
$ /opt/zcross/bin/zcross list
\end{lstlisting}

It is now possible to install \texttt{Betaboltz}:
\begin{lstlisting}[language=Bash]
$ git clone https://gitlab.com/micrenda/betaboltz.git
$ mkdir betaboltz/build && cd betaboltz/build

$ git submodule init
$ git submodule update

$ cmake .. \
  -DCMAKE_INSTALL_PREFIX=/opt/betaboltz \
  -DZCross_DIR=/opt/zcross/share/zcross/cmake
  
$ make && sudo make install
\end{lstlisting}
and the \texttt{Drifter} package:
\begin{lstlisting}[language=Bash]
$ git clone https://gitlab.com/micrenda/drifter.git
$ mkdir drifter/build && cd drifter/build

$ git submodule init
$ git submodule update

$ cmake .. \
  -DCMAKE_INSTALL_PREFIX=/opt/drifter \
  -DBetaboltz_DIR=/opt/betaboltz/share/betaboltz/cmake

$ make && sudo make install
\end{lstlisting}

The last step is to make the executable accessible from command-line, adding the new paths to the environment variables:
\begin{lstlisting}[language=Bash]
$ export PATH=$PATH:/opt/zcross/bin:/opt/drifter/bin
\end{lstlisting}

It is now possible to perform the calculation of the drift velocity using the \texttt{drifter} command-line utility:
\begin{lstlisting}[language=Bash]
$ drifter -g Ar CO2 -d 'ela:BSR/BSR-500|ine:Biagi8' 'Phelps' -x 93 7 --field 0.6 --disable-check-monoatomic

FIELD                        SYMBOL        VALUE            
run_id                                         0
events                                         1
particles                                     25
interactions                              200000
gas pressure                 P           101.325 kPa     
gas temperature              T            293.15 K       
gas density                  N       2.50348e+19 /cm³    
electric field               E               0.6 kV/cm   
electric field (red)         E/N         2.39667 Td      

drift velocity               W           4.56483 cm/μs    
                                   ±   0.0464798 cm/μs   
mobility                     μ           7608.05 cm²/V⋅s  
                                   ±     77.4664 cm²/V⋅s 
mobility (red)               μ₀          7088.99 cm²/V⋅s  
                                   ±     72.1813 cm²/V⋅s 
mobility (norm)              μN      1.90466e+23 /cm⋅V⋅s  
                                   ± 1.93935e+21 /cm⋅V⋅s 
energy                       ε          0.451974 eV       
                                   ± 4.94475e-05 eV      

townsend alpha coeff         α                 0 /cm      
                                   ±           0 /cm     
townsend alpha coeff (red)   α/N               0 cm²      
                                   ±           0 cm²            
townsend eta coeff           η                 0 /cm      
                                   ±           0 /cm     
townsend eta coeff (red)     η/N               0 cm²      
                                   ±           0 cm²     

diffusion coeff              D             12042 cm²/s    
                                   ±     10482.5 cm²/s   
diffusion coeff (norm)       ND      3.01468e+23 /cm⋅s    
                                   ± 2.62428e+23 /cm⋅s   
diffusion coeff (ratio)      D/μ         1.58279 V        
                                   ±     1.37792 V       
diffusion coeff long         Dᴸ          854.425 cm²/s    
                                   ±     1070.07 cm²/s   
diffusion coeff long (norm)  NDᴸ     2.13903e+22 /cm⋅s    
                                   ±  2.6789e+22 /cm⋅s   
diffusion coeff long (ratio) Dᴸ/μ       0.112305 V        
                                   ±    0.140655 V       
diffusion coeff tran         Dᵀ          2583.28 cm²/s    
                                   ±     2589.45 cm²/s   
diffusion coeff tran (norm)  NDᵀ     6.46718e+22 /cm⋅s    
                                   ± 6.48263e+22 /cm⋅s   
diffusion coeff tran (ratio) Dᵀ/μ       0.339546 V        
                                   ±    0.340374 V       
diffusion magnitude          σ          0.179933 cm       
                                   ±   0.0762089 cm      
diffusion magnitude long     σᴸ        0.0609852 cm       
                                   ±   0.0410937 cm      
diffusion magnitude tran     σᵀ         0.162378 cm       
                                   ±   0.0800588 cm      

real collisions                            5e+06         
null collisions                      1.34285e+07         
fail collisions                            11958         
efficiency                               27.1142 %       
elapsed time                                 264 s 
\end{lstlisting}

We want to highlight the syntax we specify the cross-section tables to use for this simulation. In the previous example, we can see we use the Phelps database \cite{phelps2019lxcat} as the $CO_2$ cross-section data. However, for the $Ar$ molecule, we specify both the BSR \cite{bsr2020lxcat} and the Biagi \cite{biagi2019lxcat} databases, separated by the vertical line. In this case, we combine the non momentum-transfer elastic cross-section tables from BSR database with the complete inelastic tables from the Biagi database, which provides exclusively momentum-transfer tables. 

Finally, in \cref{tab:commands}, we can find the list of commands used to generate the plots in this article, showing how it is possible to perform several complex analyses using basic shell scripting techniques.
\begin{table*}
\footnotesize
\begin{tabularx}{\textwidth}{l X}
	\hline 
	Figures & Command \\ 
	\hline 
	\rule[-1ex]{0pt}{3ex} \ref{fig:plot_Ar_velocity}, \ref{fig:plot_Ar_townsend}, \ref{fig:plot_Ar_diffusion}, \ref{fig:plot_Ar_energy}  & \lstinline[columns=fixed,keepspaces]{drifter -g Ar \ -d 'ela:BSR/BSR-500|ine:Biagi8' \ \ \ -e 100 -n 20 -i 250000 --reduced-field <F>} \\
	\rule[-1ex]{0pt}{3ex} \ref{fig:plot_Kr_velocity}  & \lstinline[columns=fixed,keepspaces]{drifter -g Kr \ -d 'ela:COP/Gr1|ine:Biagi8' \ \ \ \ \ \ \ -e 100 -n 20 -i 250000 --reduced-field <F>}  \\ 
	\rule[-1ex]{0pt}{3ex} \ref{fig:plot_CF4_velocity}, \ref{fig:plot_CF4_energy},  \ref{fig:plot_efficiency_gases} & \lstinline[columns=fixed,keepspaces]{drifter -g CF4 -d 'Bordage' \ \ \ \ \ \ \ \ \ \ \ \ \ \ \ \ \ \ \ \ \ \ -e 100 -n 20 -i 250000 --reduced-field <F>}  \\ 
	\rule[-1ex]{0pt}{3ex} \ref{fig:plot_CH4_velocity}, \ref{fig:plot_CH4_diffusion}, \ref{fig:plot_CH4_mobility} & \lstinline[columns=fixed,keepspaces]{drifter -g CH4 -d 'Morgan' \ \ \ \ \ \ \ \ \ \ \ \ \ \ \ \ \ \ \ \ \ \ \ -e 100 -n 20 -i 250000 --reduced-field <F>}  \\ 
	\rule[-1ex]{0pt}{3ex} \ref{fig:plot_CO2_velocity}, \ref{fig:plot_CO2_townsend}, \ref{fig:plot_CO2_diffusion},  \ref{fig:plot_efficiency_gases} & \lstinline[columns=fixed,keepspaces]{drifter -g CO2 -d 'ela:Phelps|ine:Biagi8' \ \ \ \ \ \ \ \ -e 100 -n 20 -i 250000 --reduced-field <F>}  \\ 
	\rule[-1ex]{0pt}{3ex} \ref{fig:plot_NH3_velocity} & \lstinline[columns=fixed,keepspaces]{drifter -g NH3 -d 'Morgan' \ \ \ \ \ \ \ \ \ \ \ \ \ \ \ \ \ \ \ \ \ \ \ -e 100 -n 20 -i 250000 --reduced-field <F>}  \\
	\rule[-1ex]{0pt}{3ex} \ref{fig:plot_Xe_diffusion}, \ref{fig:plot_efficiency_gases} & \lstinline[columns=fixed,keepspaces]{drifter -g Xe \ -d 'ela:cop/Gr1|ine:Biagi8' \ \ \ \ \ \ \ -e 100 -n 20 -i 250000 --reduced-field <F>}  \\
	\rule[-1ex]{0pt}{3ex} \ref{fig:plot_efficiency_zeta} & \lstinline[columns=fixed,keepspaces]{drifter -g CO2 -d 'ela:Phelps|ine:Biagi8' --reduced-field 15 --trial-algo "variable(overhead:<Z>)"} \\
	\rule[-1ex]{0pt}{3ex} \ref{fig:plot_mixture_Ar_CO2}  & \lstinline[columns=fixed,keepspaces]{drifter -g Ar CO2 -x <R1> <R2> \ \ \ \ \ \ \ \ \ \ \  --reduced-field <F>  -t 297.2} \\
	\rule[-1ex]{0pt}{3ex}  & \lstinline[columns=fixed,keepspaces]{\ \ \ \ \ \ \ \ -d 'ela.el:BSR/BSR-500|ela.mt:Biagi8|ine:Biagi8' 'ela.el:Phelps|ela.mt:Biagi8|ine:Biagi8'} \\
	\hline 
\end{tabularx} 
	\caption{List of commands used to generate the plots presented in this article. \lstinline[columns=fixed,keepspaces]{<F>,<Z>,<R1>,<R2>} represents placeholders for the reduced field [\si{\townsend}], the overhead and two gas ratios, respectively. The flag \lstinline[columns=fixed,keepspaces]{--disable-check-monoatomic} was omitted for some entries to improve readability. The cross-section tables selected here were used for our simulations, but no serious analysis was performed to determine if this section produces the best fit against experimental data.} 
	\label{tab:commands}
\end{table*}


\section{Benchmarks} \label{sec:benchmark}
A new software tool needs to demonstrate its ability to replicate data available in the literature. In this section, we will compare the swarm attributed of an electron in several molecular gases. We tested our implementation with several commonly used gases in low-temperature plasma experiments.

For benchmark, we decided to simulate \num{20} independent particles having \num{250000} real collisions. The simulation was for \num{100} times for reduced fields ranging from \SIrange{e-3}{e4}{\townsend}.

\begin{figure*}
	\centering
	\begin{subfigure}{.46\textwidth}
		\centering
		\includegraphics[width=\textwidth]{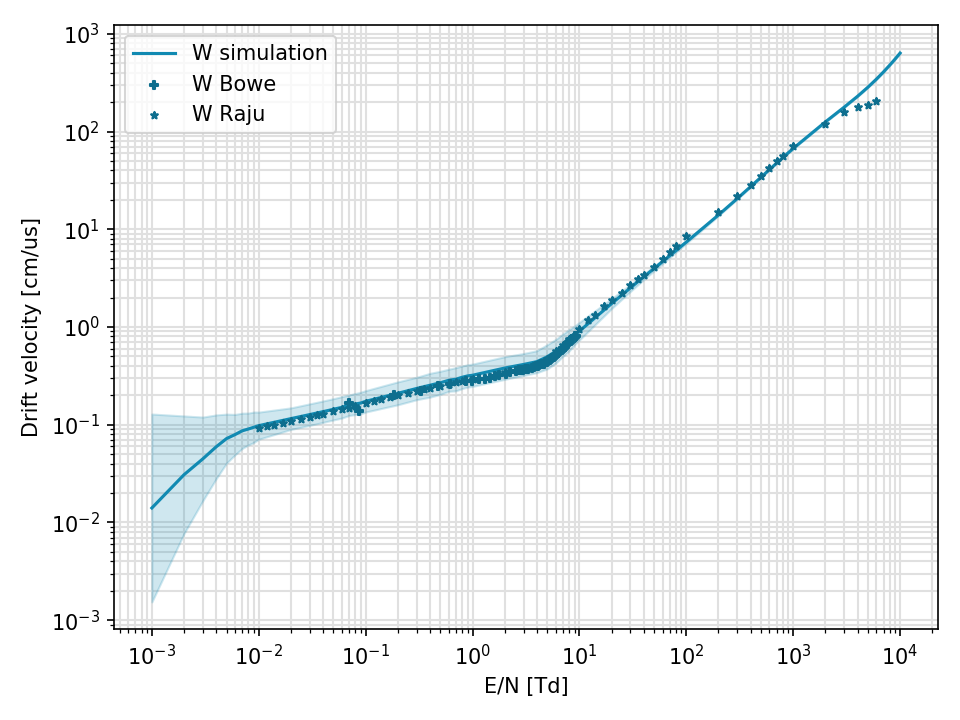}  
		\caption{Argon ($Ar$) }
		\label{fig:plot_Ar_velocity}
	\end{subfigure}
	\begin{subfigure}{.46\textwidth}
		\centering
		\includegraphics[width=\textwidth]{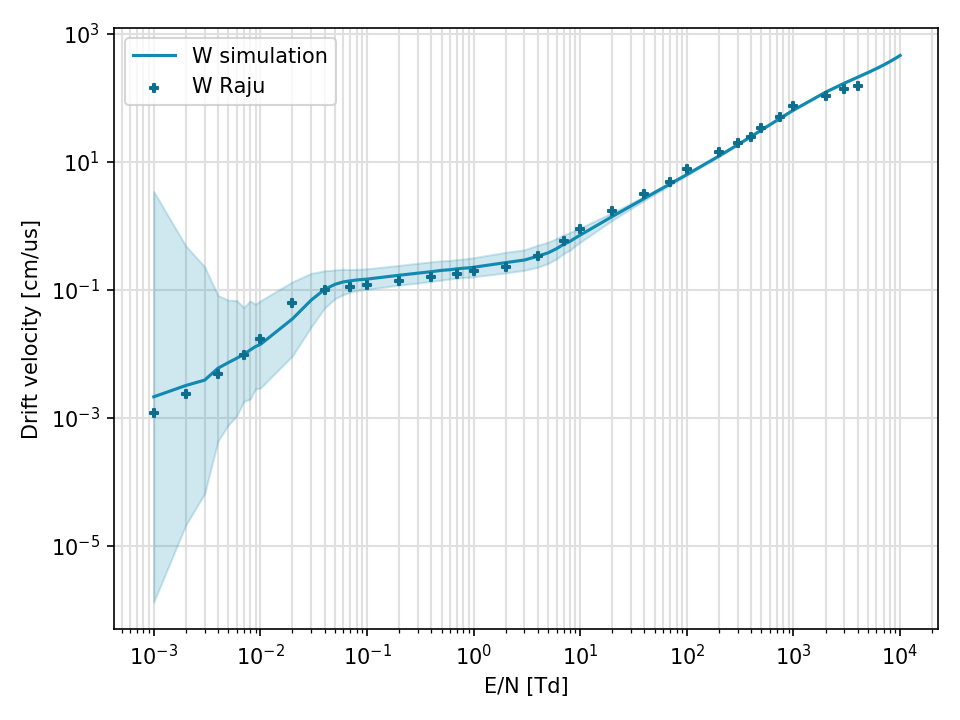}  
		\caption{Krypton ($Kr$) }
		\label{fig:plot_Kr_velocity}
	\end{subfigure}
	\begin{subfigure}{.46\textwidth}
		\centering
		\includegraphics[width=\textwidth]{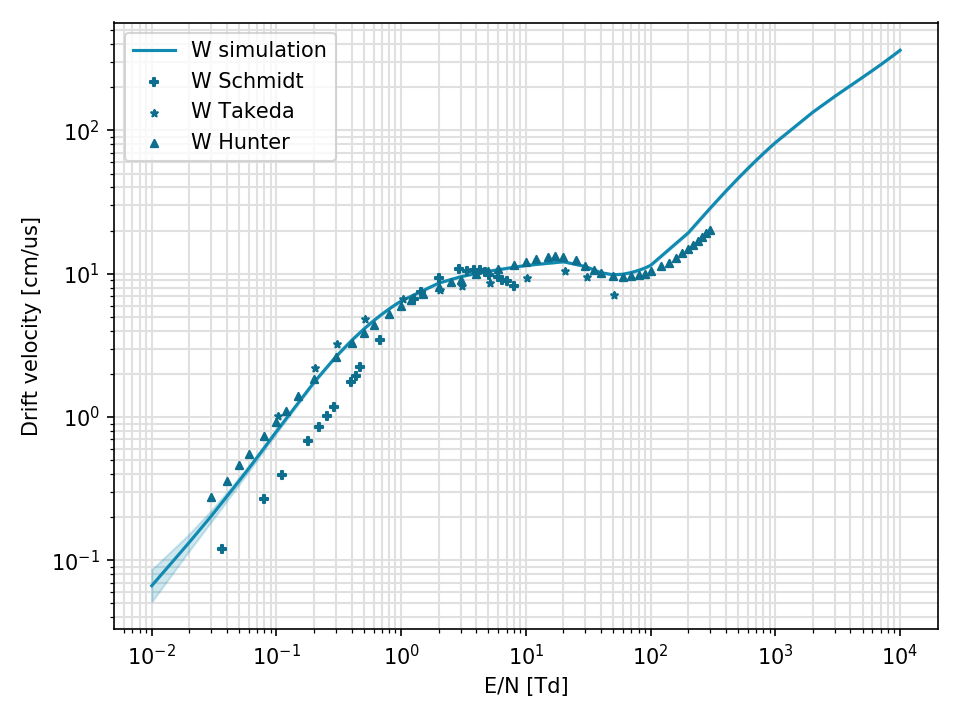}  
		\caption{Carbon tetra-fluoride ($CF_4$) }
		\label{fig:plot_CF4_velocity}
	\end{subfigure}
	\begin{subfigure}{.46\textwidth}
		\centering
		\includegraphics[width=\textwidth]{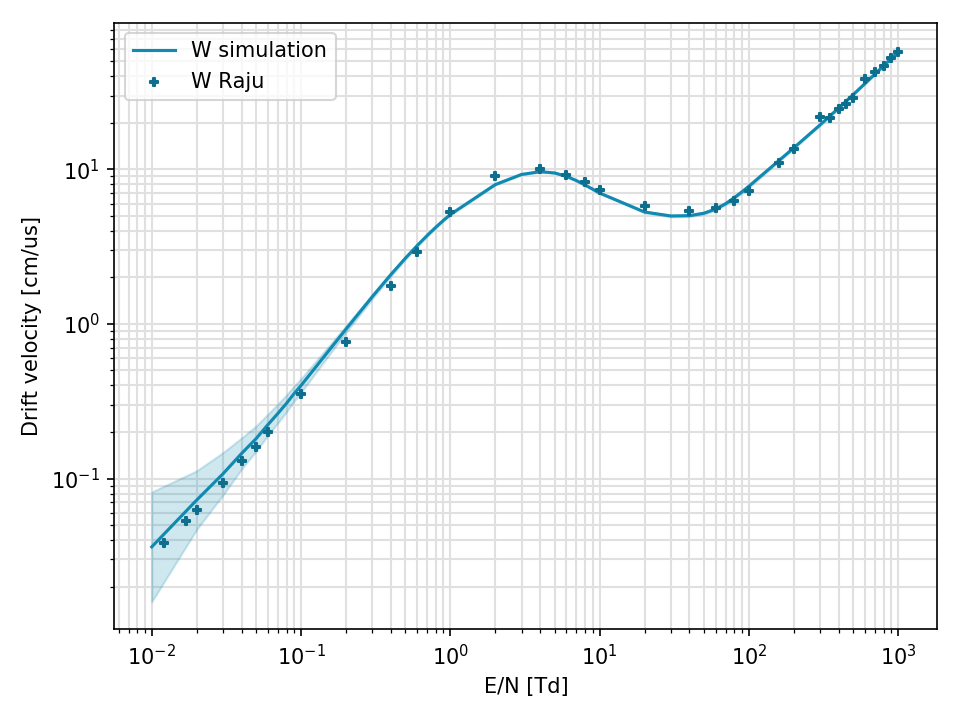}  
		\caption{Methane ($CH_4$) }
		\label{fig:plot_CH4_velocity}
	\end{subfigure}
	\begin{subfigure}{.46\textwidth}
		\centering
		\includegraphics[width=\textwidth]{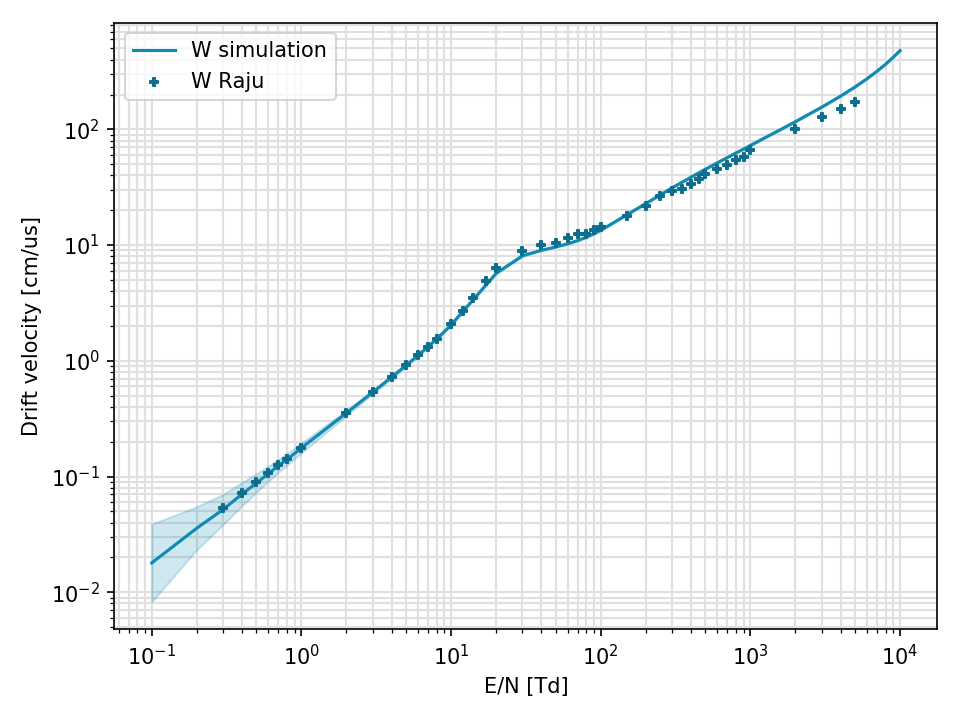}  
		\caption{Carbon Dioxide ($CO_2$) }
		\label{fig:plot_CO2_velocity}
	\end{subfigure}
	\begin{subfigure}{.46\textwidth}
		\centering
		\includegraphics[width=\textwidth]{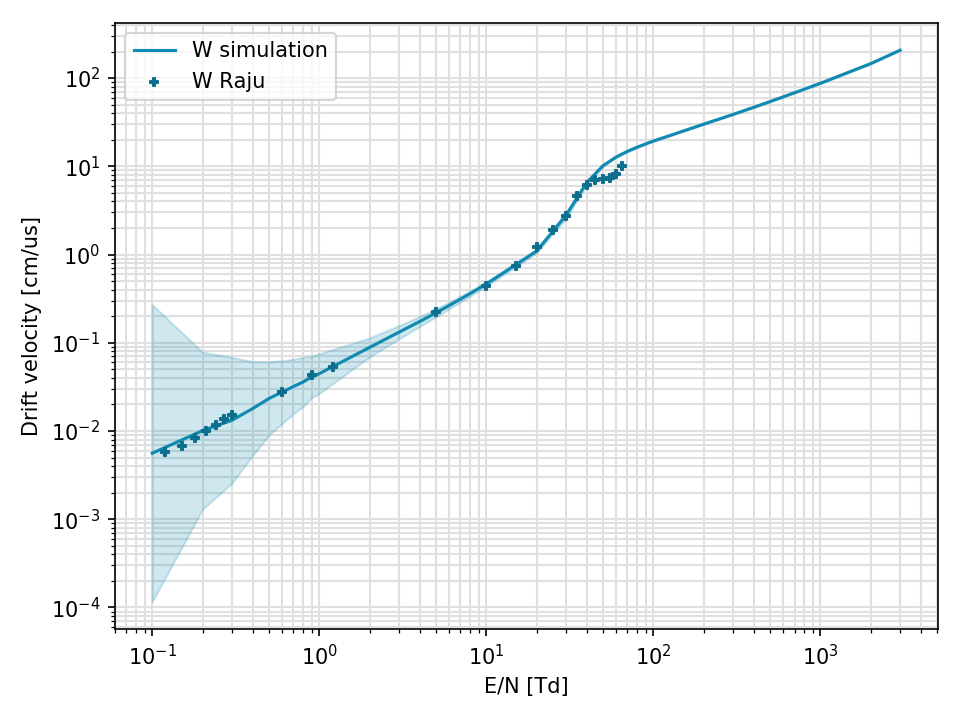}  
		\caption{Ammonia ($NH_3$) }
		\label{fig:plot_NH3_velocity}
	\end{subfigure}
	\caption{Drift velocities ($W$) simulated for different gaseous molecules: \SI{200}{} particles having \SI{250000}{} real collisions each. Bands represent standard deviations. Cross-section tables taken from \cite{biagi2019lxcat,bsr2020lxcat,phelps2019lxcat,bordage2019lxcat,morgan2019lxcat}. Reference data taken from \cite{bowe1960drift,schmidt1988electron,takeda1993mobility,hunter1988electron,raju_gaseous_2011}.}
	\label{fig:plot_velocity}
\end{figure*}

The plots presented in \cref{fig:plot_velocity} are in good agreement with reference data. However, we can observe, in last points of the reference data, the simulation tends to over-estimate the drift velocities (this effect is clearly visible in \cref{fig:plot_Ar_velocity} for an electric field grater than \SI{3000}{\townsend}). We do not have a clear explanation for this, but we suspect we are ignoring a dissipative process such as a multiple ionization reactions or radiation emission by charged particles.

\begin{figure*}
	\centering
	\begin{subfigure}{.46\textwidth}
		\centering
		\includegraphics[width=\textwidth]{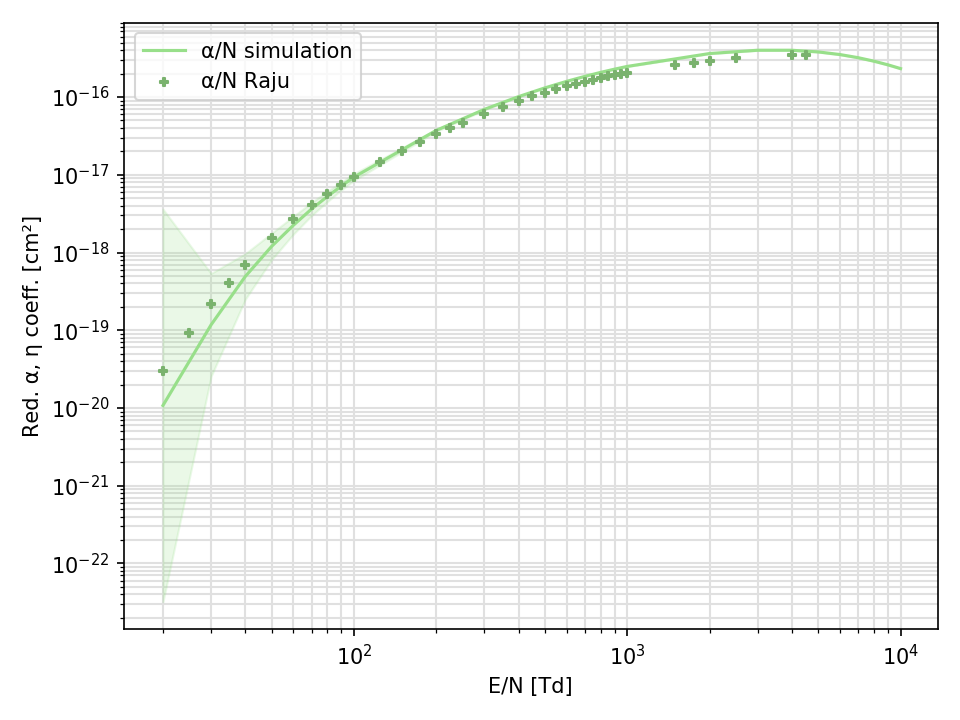}  
		\caption{Argon ($Ar$) }
		\label{fig:plot_Ar_townsend}
	\end{subfigure}
	\begin{subfigure}{.46\textwidth}
		\centering
		\includegraphics[width=\textwidth]{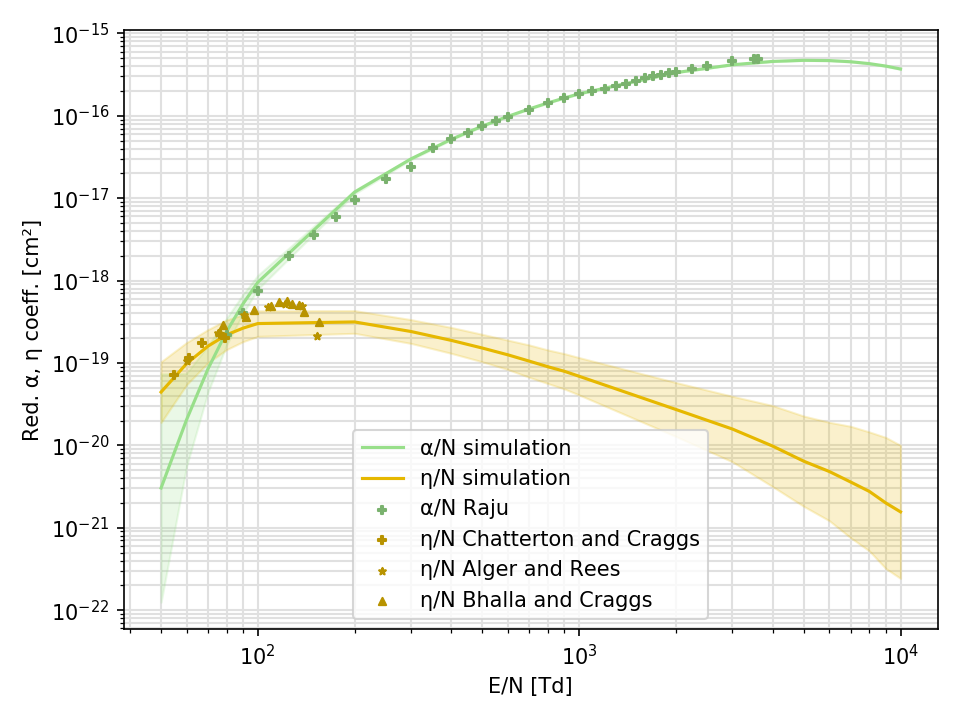}  
		\caption{Carbon Dioxide ($CO_2$) }
		\label{fig:plot_CO2_townsend}
	\end{subfigure}
	\caption{Reduced first ionization $\alpha / N$ and attachment $\eta / N$ coefficients for $Ar$ and $CO_2$: \SI{200}{} particles having \SI{250000}{} real collisions each. Bands represent standard deviations. Cross-section tables taken from \cite{biagi2019lxcat,bsr2020lxcat,phelps2019lxcat}. Reference data taken from \cite{bhalla1960measurement,chatterton1965attachment,alger1976ionization,raju_gaseous_2011}.}
	\label{fig:plot_townsend}
\end{figure*}

In \cref{fig:plot_townsend}, we can see the first reduced ionization and attachment coefficients for an atomic and molecular gas. We can notice, in this case, we tend to underestimate the value of the $\alpha$ coefficient for low fields, near the ionization threshold energy, linked with an increased standard deviation.

\begin{figure*}
	\centering
	\begin{subfigure}{.46\textwidth}
		\centering
		\includegraphics[width=\textwidth]{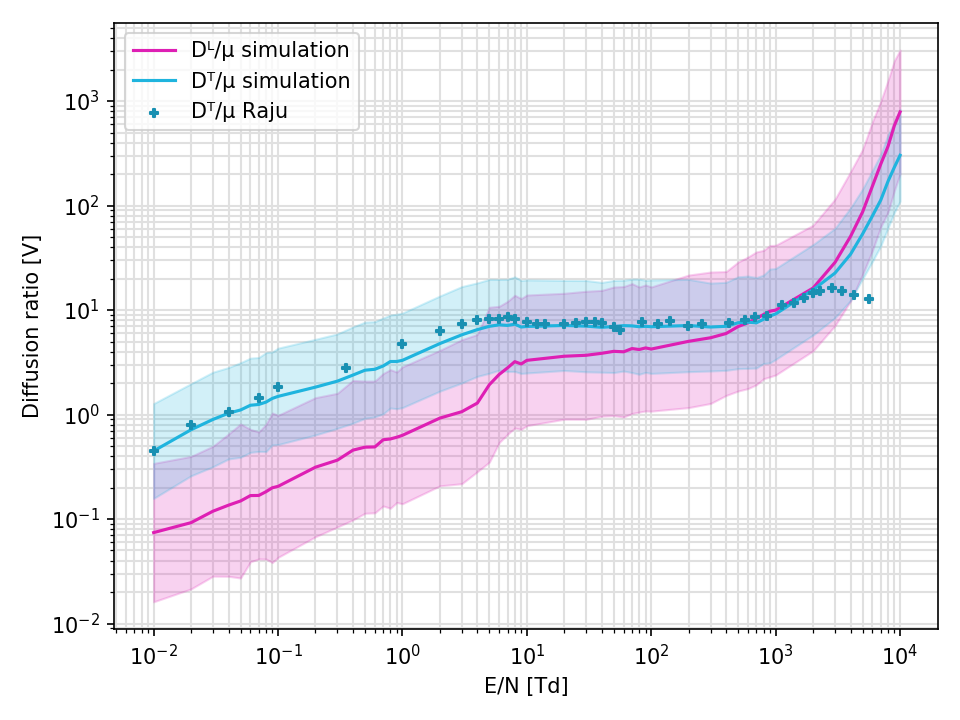}  
		\caption{Argon ($Ar$) }
		\label{fig:plot_Ar_diffusion}
	\end{subfigure}
	\begin{subfigure}{.46\textwidth}
		\centering
		\includegraphics[width=\textwidth]{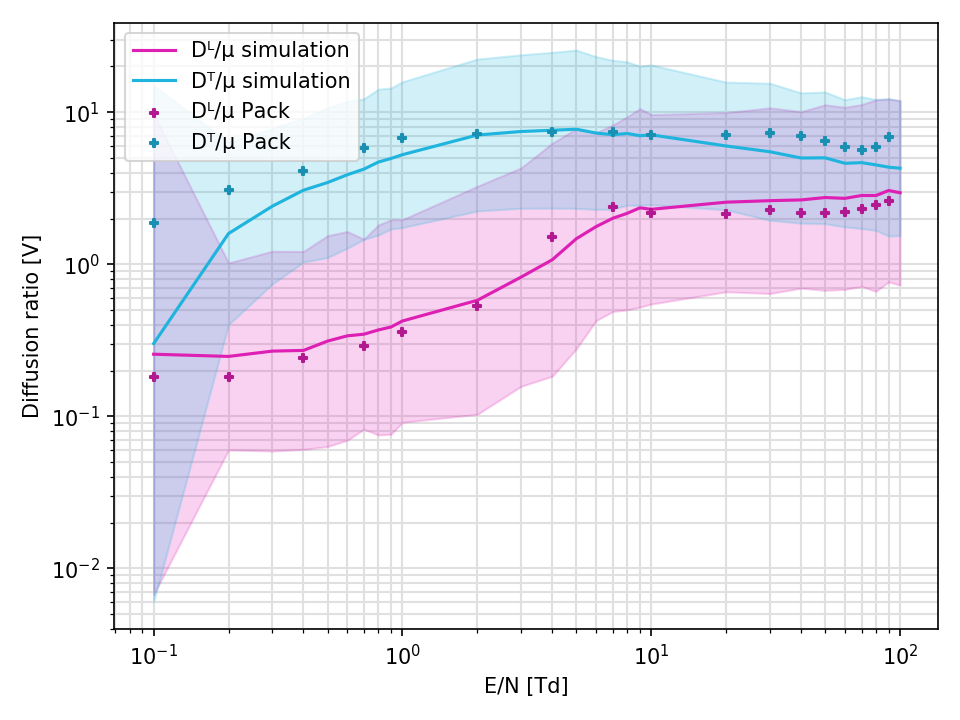}  
		\caption{Xenon ($Xe$) }
		\label{fig:plot_Xe_diffusion}
	\end{subfigure}
	\begin{subfigure}{.46\textwidth}
		\centering
		\includegraphics[width=\textwidth]{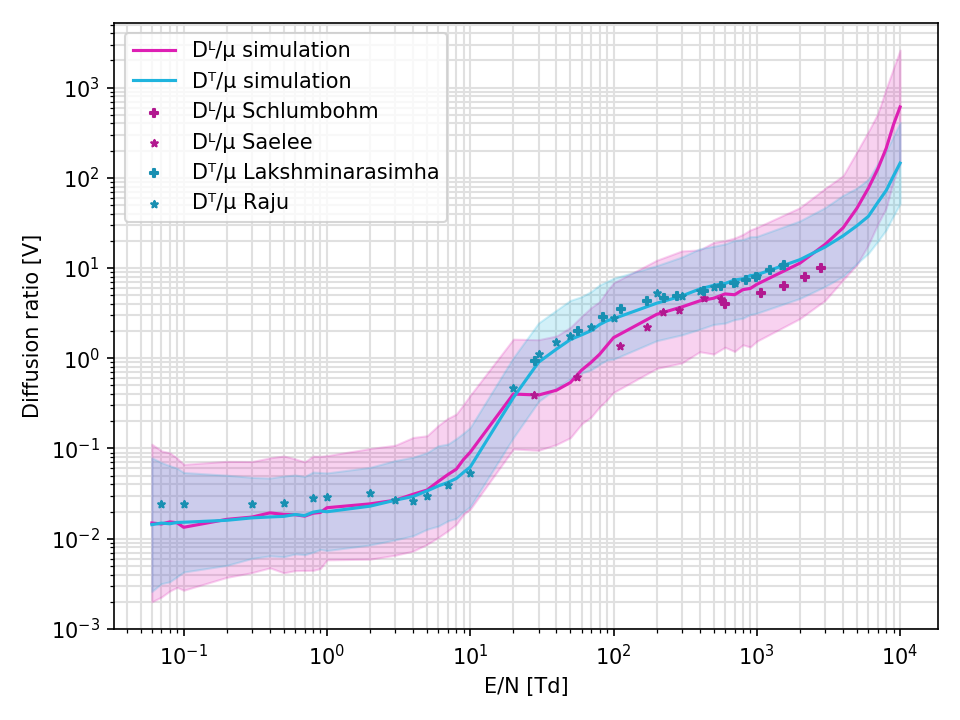}  
		\caption{Carbon Dioxide ($CO_2$) }
		\label{fig:plot_CO2_diffusion}
	\end{subfigure}
	\begin{subfigure}{.46\textwidth}
		\centering
		\includegraphics[width=\textwidth]{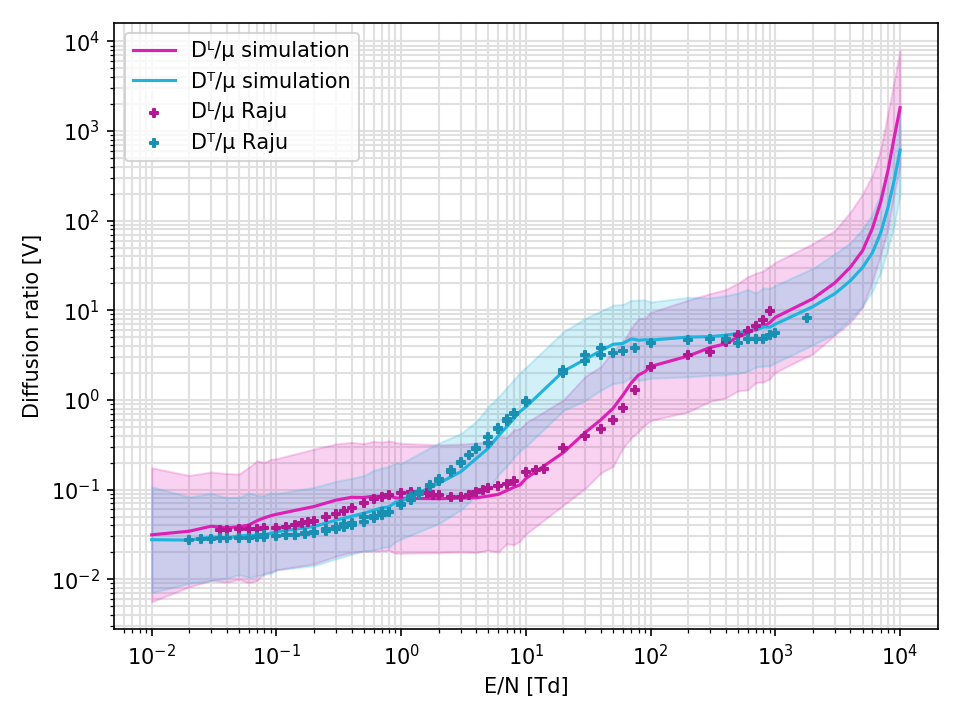}  
		\caption{Methane ($CH_4$) }
		\label{fig:plot_CH4_diffusion}
	\end{subfigure}
	\caption{Transversal ($D^T/\mu$) and longitudinal ($D^L\mu$) diffusion coefficients for $Xe$, $CH_4$ and $CO_2$: \SI{200}{} particles having \SI{250000}{} real collisions each. Bands represent standard deviations. Cross-section tables taken from \cite{biagi2019lxcat,phelps2019lxcat,cop2020lxcat,morgan2019lxcat}. Reference data taken from \cite{pack1992longitudinal,kucukarpaci1979simulation,raju_gaseous_2011}.}
	\label{fig:plot_diffusion}
\end{figure*}

We present, in \cref{fig:plot_diffusion}, a calculation of longitudinal and transversal diffusion. This task was the most challenging because, due to how it is calculated, we need to simulate $\approx \num{1e9}$ interactions for each field value to get reasonable standard deviations, especially at low fields values. We can see a reasonable match with data available in the literature: however, for fields lower than \SI{1}{\townsend}, we have a mediocre match and unsatisfactory standard deviations. Simulations with an increased number of interactions may improve the match and reduce the data dispersion.

\begin{figure*}
	\centering
	\begin{subfigure}{.46\textwidth}
		\centering
		\includegraphics[width=\textwidth]{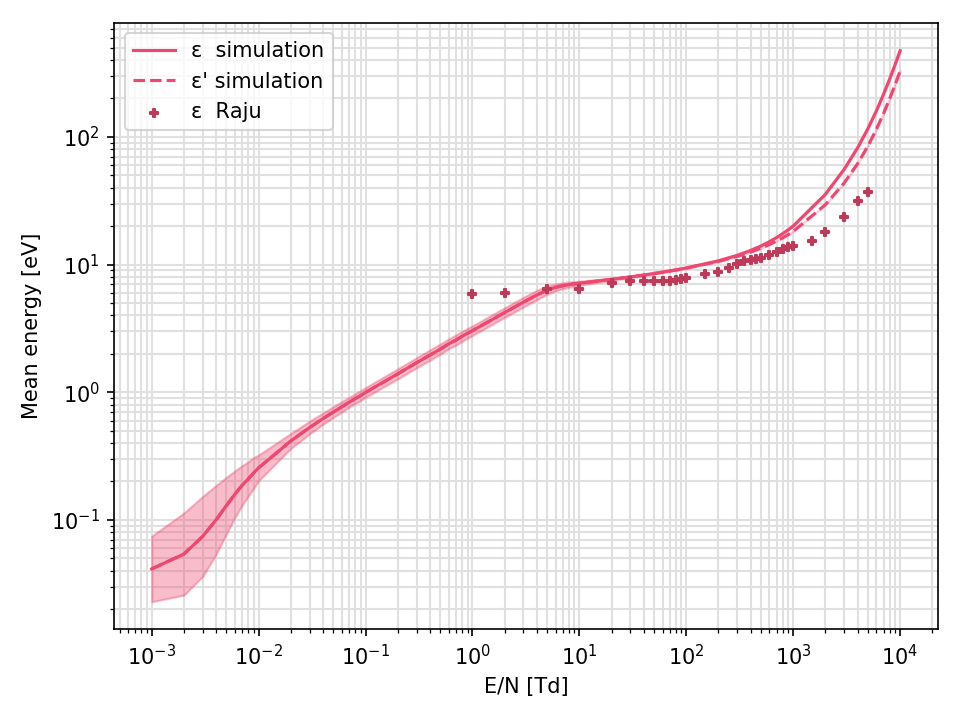}  
		\caption{Argon ($Ar$) }
		\label{fig:plot_Ar_energy}
	\end{subfigure}
	\begin{subfigure}{.46\textwidth}
		\centering
		\includegraphics[width=\textwidth]{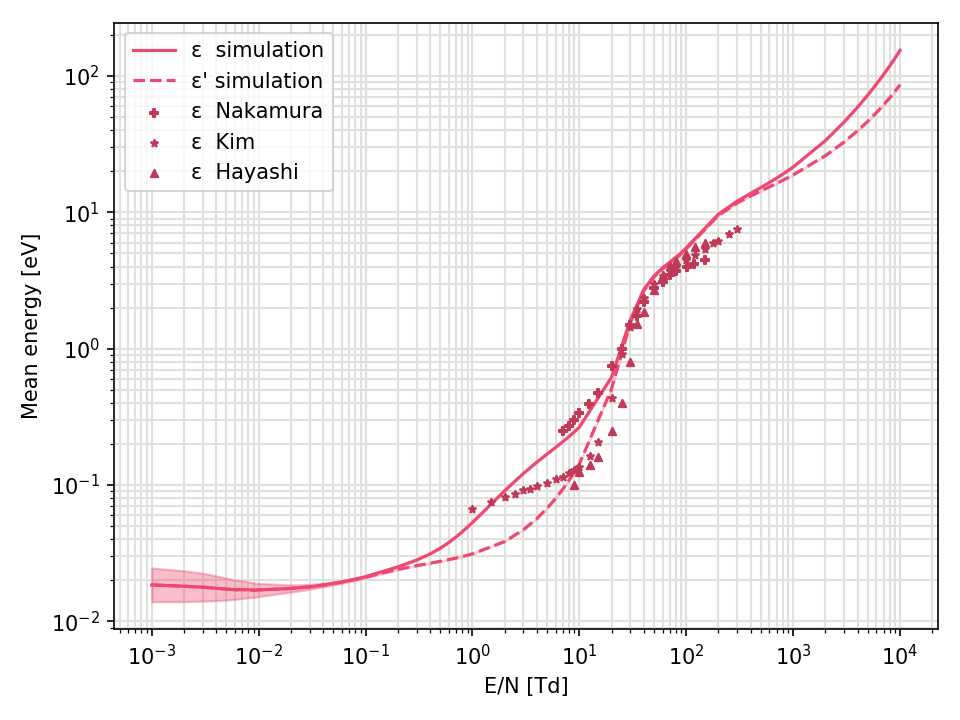}  
		\caption{Carbon tetra-fluoride ($CF_4$) }
		\label{fig:plot_CF4_energy}
	\end{subfigure}
	\caption{Mean electron energy sampled just before ($\varepsilon$) and just after ($\varepsilon'$) the collision: \SI{200}{} particles having \SI{250000}{} real collisions each. Bands represent standard deviations. Cross-section tables taken from \cite{biagi2019lxcat,bsr2020lxcat,bordage2019lxcat}. Reference data taken from \cite{kim2015electron,raju_gaseous_2011}.}
	\label{fig:plot_energy}
\end{figure*}

\begin{figure*}
	\centering
	\begin{subfigure}{.46\textwidth}
		\centering
		\includegraphics[width=\columnwidth]{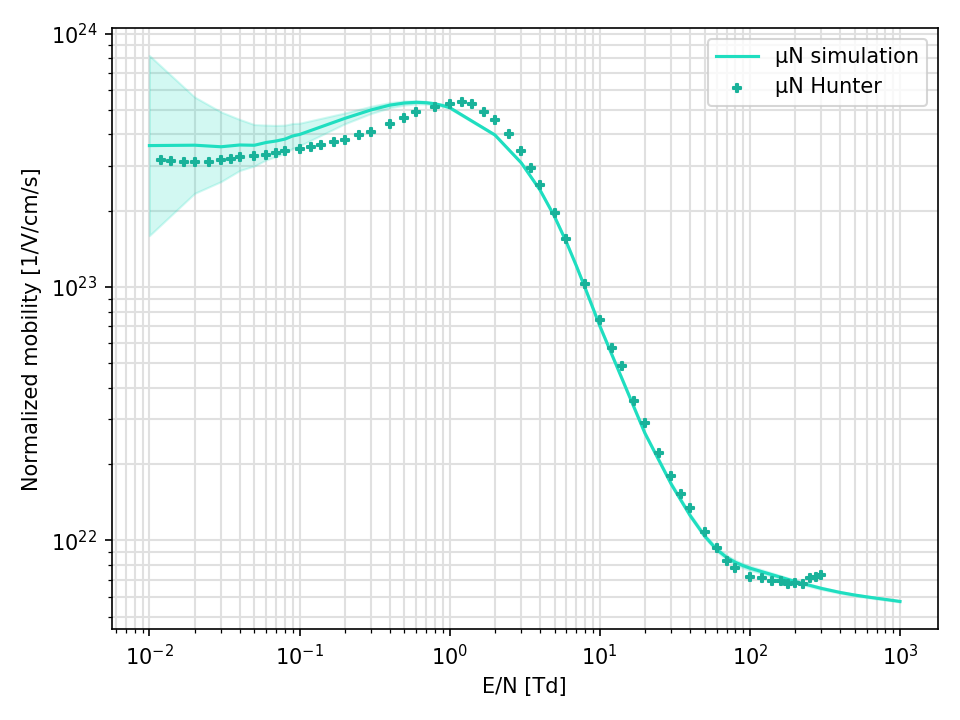}
		\caption{Methane ($CH_4$) }
		\label{fig:plot_CH4_mobility}
	\end{subfigure}  
	\caption{Normalized mobility ($\mu N$): \SI{200}{} particles having \SI{250000}{} real collisions each. Bands represent standard deviations. Cross-section tables taken from \cite{morgan2019lxcat}. Reference data taken from \cite{hunter1986electron}.}
	\label{fig:plot_mobility}
\end{figure*}

In \cref{fig:plot_energy,fig:plot_mobility}, we present the simulated mean energy of the electrons and their mobilities. The energy is sampled immediately before, $\varepsilon$, and after, $\varepsilon$', the collision, and its mean values (and standard deviation) are reported. We can observe, as expected, the energy loss, $\varepsilon$-$\varepsilon$', increases at high field values when inelastic processes become predominant. We can observe, in \cref{fig:plot_Ar_energy,fig:plot_CF4_energy}, we have a not-so-good match with experimental data. At the time, we are unsure about the root cause of this mismatch.
 
\begin{figure}
	\includegraphics[width=\columnwidth]{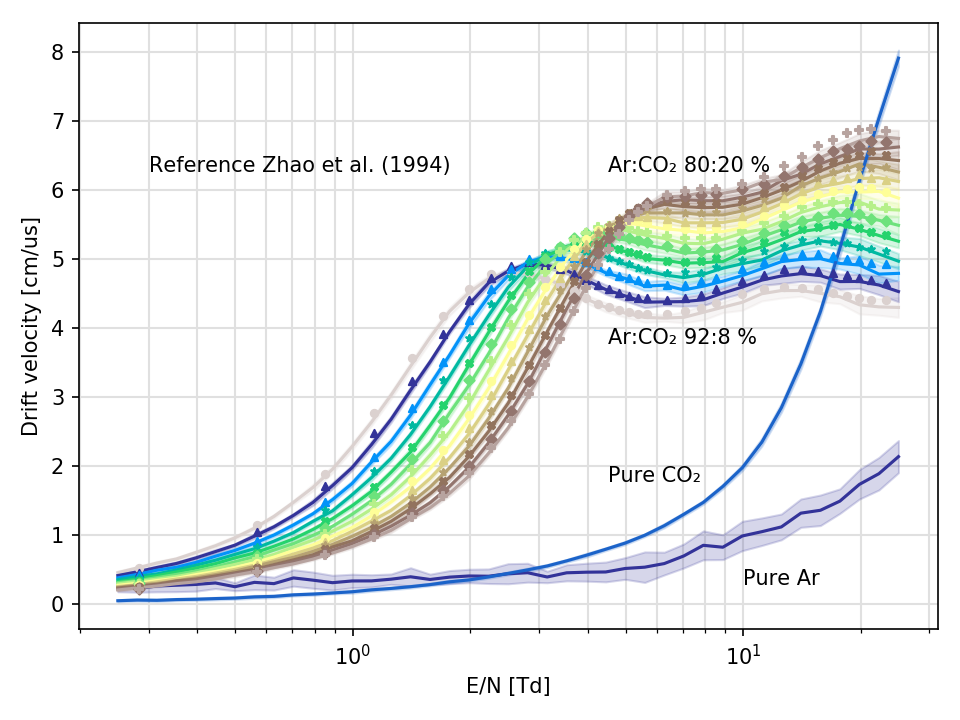}
	\caption[Drift velocity in $Ar:CO_2$]{Drift velocity ($W$) for $Ar:CO_2$ mixture between \SIrange{0.25}{25}{\townsend} at \SI{24}{\celsius} and \SI{101.325}{\kilo\pascal}. For each gas-ratio and electric field where simulated \num{25} particles with \num{200000} real interactions each. Bands represent standard deviations. Cross-section data taken from \cite{bsr2020lxcat,biagi2019lxcat,phelps2019lxcat}. Reference data taken from \cite{zhao1994study}.}
	\label{fig:plot_mixture_Ar_CO2}
\end{figure}

Finally, in \cref{fig:plot_mixture_Ar_CO2}, we decided to test our implementation in a custom $Ar-CO_2$ gas mixture. As reference data, we decided to use \cite{zhao1994study}. We have a good match over the entire energy range: however, for field higher than \SI{10}{\townsend}, we tend to underestimate the drift velocity by a factor of $\approx$ \SI{15}{\percent}.

To evaluate the performance of the multi-thread implementation, we decided to perform the simulation of an electron under a field of, respectively, \SIlist{1;5;40}{\kilo\volt\per\centi\meter} in a $Ar:CO_2$ $93:7$ mixture at standard conditions. To remove any influence from the detector geometry, we decide to place the seed electron in an infinite volume and track the movement for \SI{1}{\nano\second}. The electric fields were chosen to highlight the different performance gain existing in drift only (\SIlist{1;5}{\kilo\volt\per\centi\meter}) and in the avalanche regime (\SI{40}{\kilo\volt\per\centi\meter}).

\begin{figure}
	\includegraphics[width=\columnwidth]{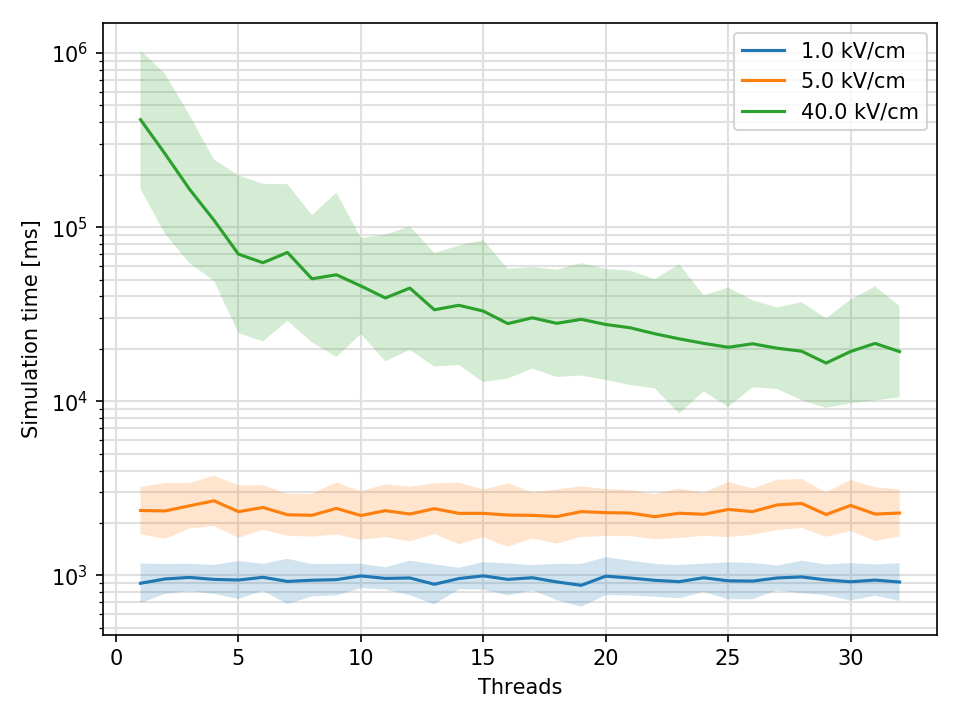}
	\caption[Performance of multi-thread simulation]{Computing time for the \SI{1}{\nano\second} simulation of a single electron avalanche under a \textit{z-}aligned static electric field in $Ar:CO_2$ at $93:7$. The simulation was executed \SI{50}{} times for each setting and the average values were displayed. Bands represent standard deviation. The simulation was executed on a \num{288} threads Intel(R) Xeon Phi CPU 7290 at \SI{1.50}{\giga\hertz} with \SI{188}{\gibi\byte} of RAM.}
	\label{fig:plot_performance}
\end{figure}

In \cref{fig:plot_performance}, it is possible to see how there is no performance gain for drift only regime: this behavior can be explained due to the nature of the collision processes happening at lower energy, mainly elastic and inelastic collisions, which can not be executed concurrently.
At higher field values, electrons gain enough energy to enable ionization processes: in this regime, it is possible to increase the simulation performance using additional threads, transforming it into an \textit{embarrassingly parallel} problem \cite{herlihy2012art}. In the latter case, we can achieve a performance gain proportional to the number of running threads available for the simulation.

\section{Limitations and future plans} \label{sec:limitation}

In this section, we will discuss the current limitations of the presented library and future development plans. The most impelling limitation regards the model we are using to determine the scattering angle in the collision, as shown in \cref{eq:collision_angle}. We intend to refine the model to cover any gaseous molecule, including the polar ones.

We decided to assume that the electron density is always sufficiently small to ignore Coulomb interactions  between the particles. We realize this is a strong assumption in some conditions (e.g., when having strong avalanche amplifications): however, this assumption simplified the multi-thread implementation of the presented solution.   

In addition, we decide to ignore the penning ionization at this stage. Some required data to evaluate this effect are already present in the \texttt{ZCross} interface (such as the threshold energy for each inelastic process). However, some other parameters, such as the relaxation time of the excited states, are not yet available. We are evaluating the possibility of adding such feature in a future implementation.

During the program’s validation, we noticed the simulation of oxygen-based gas mixtures has poor results at lower energy range: molecular $O_2$ contains some resonances in the range \SIrange{0.1}{1}{\electronvolt}, which make difficult the simulation using the \textit{null-collisions} technique, leading to incorrect results.

Similar problems were found during the simulation of $Ar-CF_4$ gas mixtures: while the simulation of pure $CF_4$ lead, somehow, to reasonable results, its behavior in mixture leads to poor results. Moreover, the drift velocities are strongly dependent on the cross-section database used. This may be caused by an insufficient coverage of the cross-section tables of the used database. As a design decision, we do not offer any set of recommended tables and the final user is fully responsible for the selection of the cross-section data. However, we feel that in the future, we will have to provide a solution to let the final user to objectively evaluate the cross-section databases available.


Because, for the moment, this code can only simulate uniform static fields, we intend to include a cylindrical electromagnetic field feature. It is also foreseen the implementation of the support for arbitrary fields, not necessarily time-independent, enabling the simulation of more complex detectors.

Finally, to improve the final user experience and offer better support, we decided to create a public forum dedicated to the low-temperature plasma community. Here, researchers and students can receive support and provide feedback about the software presented in this article. The forum can be accessed at the address:
\begin{center}
	\url{https://forum.cold-plasma.org}
\end{center}


\section{Conclusions} \label{sec:conclusion}

We developed and tested a \texttt{C++} shared library with multi-thread and dimensional static analysis support, for the simulation of electron and ion transport in arbitrary gas mixtures with static uniform electric and magnetic fields. We presented a short theoretical introduction, and we have described in detail our open-source code architecture.

The framework was validated against several available data, and we found a good agreement for atomic gases and molecular gases for low and middle range reduced fields.

Future development plans include implementing a refined model to better handle linear and polar molecules, adding the support for non-uniform and non-static fields.

This software is distributed under \textit{LGPL v3} license and is free for any use, including the linking into proprietary software under the terms of the license. No expressed or implied warranty is provided with this software, and the end-user is responsible for the correctness of the results provided by this library.

\section*{Acknowledgments}
We would like to specially thanks Prof. Siu A. Chin, for the kind support given during the implementation of the relativistic formula, Prof. Theodoros Alexopoulos, for valuable input on the simulation of MicroMegas detectors, Prof. C\u{a}lin Alexa  for his valuable suggestions, guidance and for believing on this project since the beginning. His determination and its unconditioned support were the soul of this article. We want to thank our IFIN-HH ATLAS group colleagues, for the constructively and friendly environment they create every day.

A grateful thought goes to Gorur G. Raju, for his book \cite{raju_gaseous_2011} which was fundamental during the validation of this project: we think the amount of work required to collect and review the countless tables in his book is inferior only to the benefit he gave to the whole low-pressure plasma community. A final recognition goes to the anonymous reviewers of this article, whose valuable feedback has pushed us to resolve some critical issues in the previous revisions of this document. 

We would also like to thank the LXCat team for their support and for providing hospitality for the custom cross-section XML format in the LXCat site.

This work was supported by the research grants \texttt{ATLAS CERN-RO} and \texttt{PN19060104}.
	
	
	
	\bibliographystyle{elsarticle-num}
	\bibliography{article}

\end{document}